\documentclass[twocolumn,superscriptaddress]{revtex4}
\usepackage{graphicx}
\usepackage{amsmath}

\newcommand{\remove}[1]{}

\def\be{\begin{equation}}
\def\ee{\end{equation}}
\def\ba{\begin{eqnarray}}
\def\ea{\end{eqnarray}}

\usepackage{placeins}

\begin{document}


\title{Status of the GRANIT facility}

\author{Damien Roulier}
\email{roulier@ill.fr}
\affiliation{Universit\'e Grenoble-Alpes, 38000 Grenoble, France}
\affiliation{Institut Max von Laue - Paul Langevin, 71 av. des Martyrs, 38000 Grenoble, France}

\author{Francis Vezzu}
\affiliation{LPSC Grenoble, Universit\'e Grenoble Alpes, CNRS/IN2P3, 53 av. des Martyrs, 38026 Grenoble Cedex, France}

\author{Stefan Baessler}
\affiliation{Physics Department, University of Virginia 382 McCormick Road, Charlottesville, VA 22904, U.S.A}
\author{Beno\^it Cl\'ement}
\affiliation{LPSC Grenoble, Universit\'e Grenoble Alpes, CNRS/IN2P3, 53 av. des Martyrs, 38026 Grenoble Cedex, France}
\author{Daniel Morton}
\affiliation{Physics Department, University of Virginia 382 McCormick Road, Charlottesville, VA 22904, U.S.A}
\author{Valery Nesvizhevsky}
\affiliation{Institut Max von Laue - Paul Langevin, 71 av. des Martyrs, 38000 Grenoble, France}
\author{Guillaume Pignol}
\author{Dominique Rebreyend}
\affiliation{LPSC Grenoble, Universit\'e Grenoble Alpes, CNRS/IN2P3, 53 av. des Martyrs, 38026 Grenoble Cedex, France}

\date{\today}


\begin{abstract}
The GRANIT facility is a follow-up project, which is motivated by the recent discovery of gravitational quantum states of ultracold neutrons. The goal of the project is to approach the ultimate accuracy in measuring parameters of such quantum states and also to apply this phenomenon and related experimental techniques to a broad range of applications in particle physics as well as in surface and nanoscience studies. We overview the current status of this facility, the recent test measurements and the nearest prospects.
\end{abstract}

\maketitle


\section{Introduction}

The GRANIT facility~\cite{Baessler2011707,SchmidtWellenburg2009267} is a follow-up project, which is motivated by the recent discovery of gravitational quantum states of ultracold neutrons (UCNs)~\cite{NesvizhevskyNature,PhysRevD.67.102002,NesvizhevskyEP}. The main goal of the project is to realize the resonance spectroscopy of those quantum states, with the prospect of achieving an unprecedented sensitivity. Such precision measurements would address in particular searches for extra short-range fundamental forces~\cite{PhysRevD.75.075006,PhysRevD.77.034020,Antoniadis2011755,PhysRevLett.107.111301}. Also the phenomenon of gravitational quantum states and related experimental techniques could be applied to a broad range of other applications in particle physics as well as in surface and nanoscience studies~\cite{Antoniadis2011703}.

\begin{figure}

\begin{center}
\includegraphics[width=0.49\textwidth]{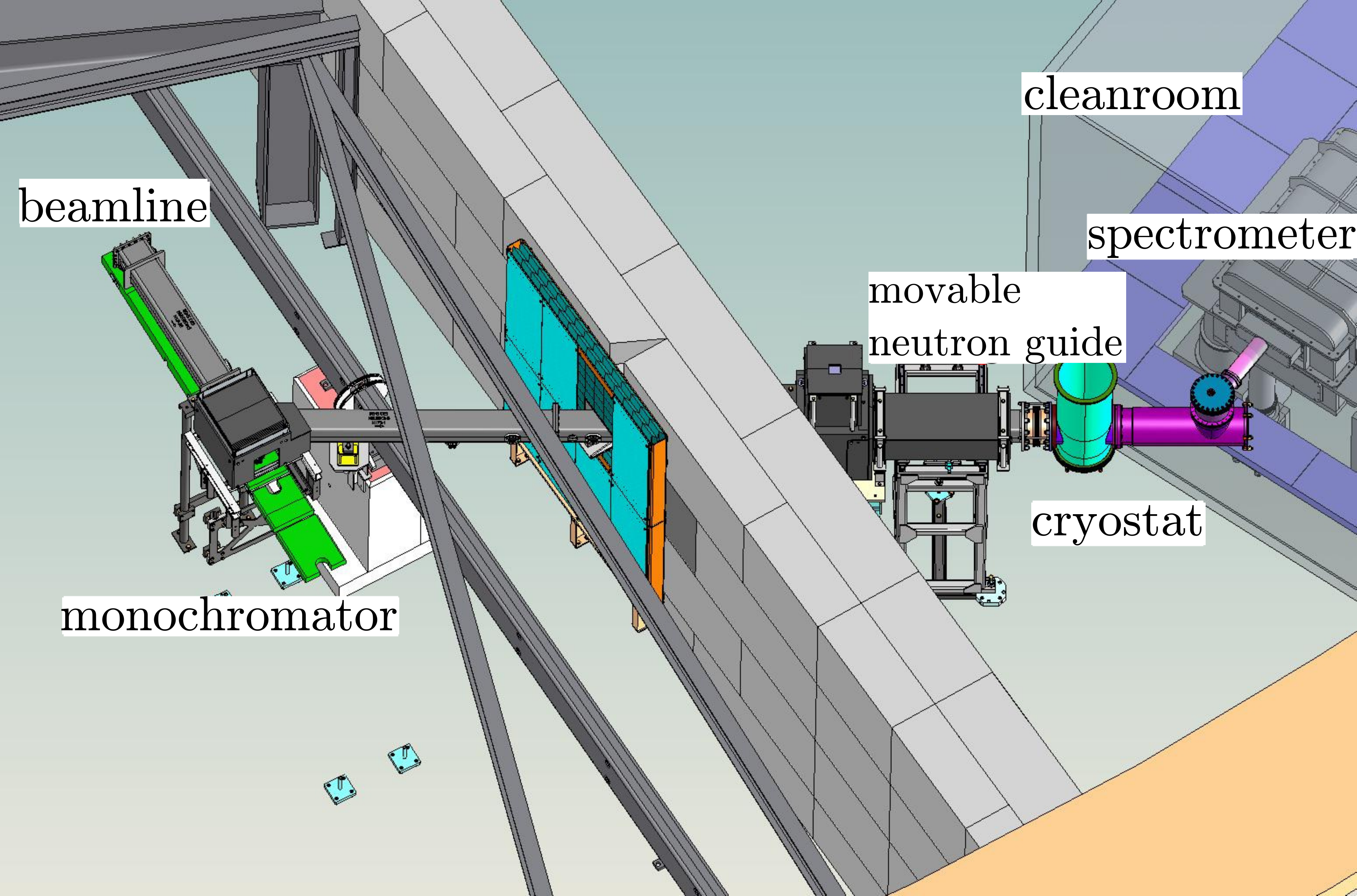}
\end{center}
\caption{The GRANIT instrument at Level C of ILL, Grenoble.}
\label{granit_location}
\end{figure}

\begin{figure}

\begin{center}
\includegraphics[width=0.49\textwidth]{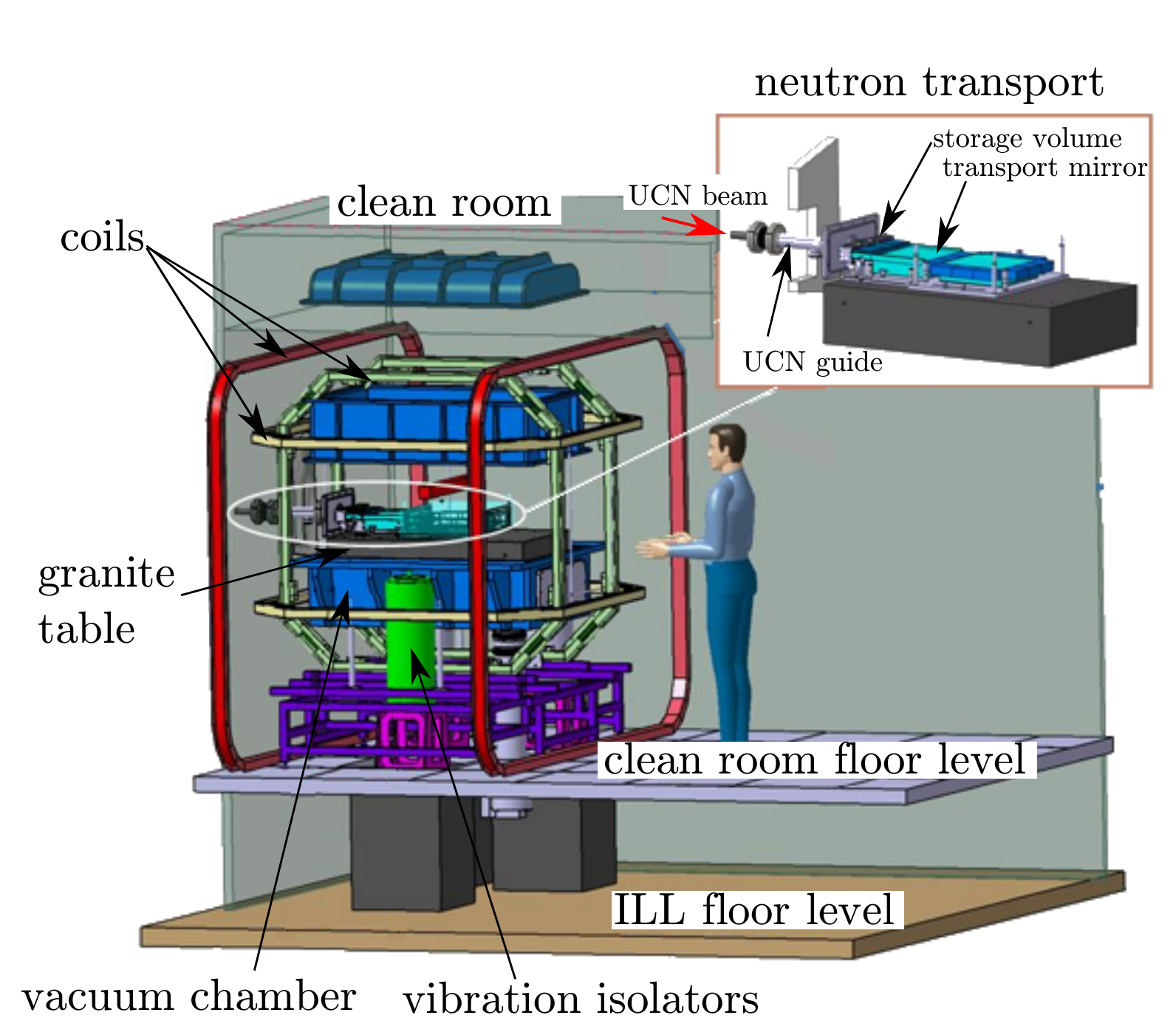}
\end{center}
\caption{The spectrometer in the cleanroom. Measurement and control instruments are installed on a flat massive granite table with the mass of $550$ kg, all inside a welded aluminum vacuum chamber with passivated wall surfaces. All this is placed in a clean controlled area (ISO 5) in order to protect the delicate optical elements.}
\label{granit_spectro}
\end{figure}

GRANIT is located at the level C of the Institut Laue-Langevin (ILL) in Grenoble, at the H172A beamline as shown in Fig.~\ref{granit_location}. The instrument comprises an ultracold neutron source based on the production of UCNs in superfluid helium and a spectrometer installed in an ISO 5 class clean room, pictured in Fig.~\ref{granit_spectro}. A monochromatic neutron beam ($0.89$~nm wavelength) is extracted from a white cold neutron beam with a monochromator~\cite{Courtois2011S37} and guided towards the superfluid helium bath where UCNs are produced~\cite{SchmidtWellenburg2009267,PhysRevLett.107.134801,Piegsa}. An extraction guide allows then to transfer those UCNs to the spectrometer. 

In the spectrometer, UCNs are first stored in an intermediate storage volume. To exit this volume, neutrons must go through an extraction slit of height $\approx 100$~$\mu$m, a compromise between the total UCN flux and the UCN phase-space density. Then neutrons will bounce over high quality mirrors very close to the surface and the method of resonance spectroscopy~\cite{NesvizhNY} will be applied. The resonance can be induced by a vibration of the bottom mirror, as used by the QBounce collaboration~\cite{Jenke} or an oscillating magnetic field gradient as in the GRANIT spectrometer~\cite{PignolTR}.

In 2013 we have performed extensive tests of the various components of the facility and connected for the first time the source to the spectrometer. In this article we will present the characterization of the whole UCN chain: the $0.89$~nm neutron beam, the cryogenic production volume, the extraction guides, and the mirror assembly. We also present a confrontation of the measurements to Monte-Carlo simulations. Finally we present the first UCN flux measurement in the GRANIT spectrometer.

\section{$\bf 0.89~\rm \bf nm$ neutrons beam}

In superfluid helium, cold neutrons with the wavelength of $\lambda = 0.89$~nm can be converted into UCNs through resonant phonon excitation~\cite{Golub1977337}. Therefore, the UCN production rate in the source will depend directly on the neutron flux at this precise wavelength $\left.\frac{d\Phi}{d\lambda}\right|_{0.89~\rm nm}$.

\subsection{Monochromator adjustment}

The monochromator~\cite{Courtois2011S37} is composed of 18 intercalated stage-2 KC$_{24}$ crystals, with a lattice constant of $d=8.74$~\AA.
According to the Bragg formula for the first order reflection, $2d \sin \theta = \lambda$, the outgoing beamline angle corresponding to $\lambda = 0.89$~nm is found to be $2 \theta = 61.2$~degrees, defining the geometry of the installation downstream.

\begin{figure}
\begin{center}
\includegraphics[width=0.49\textwidth]{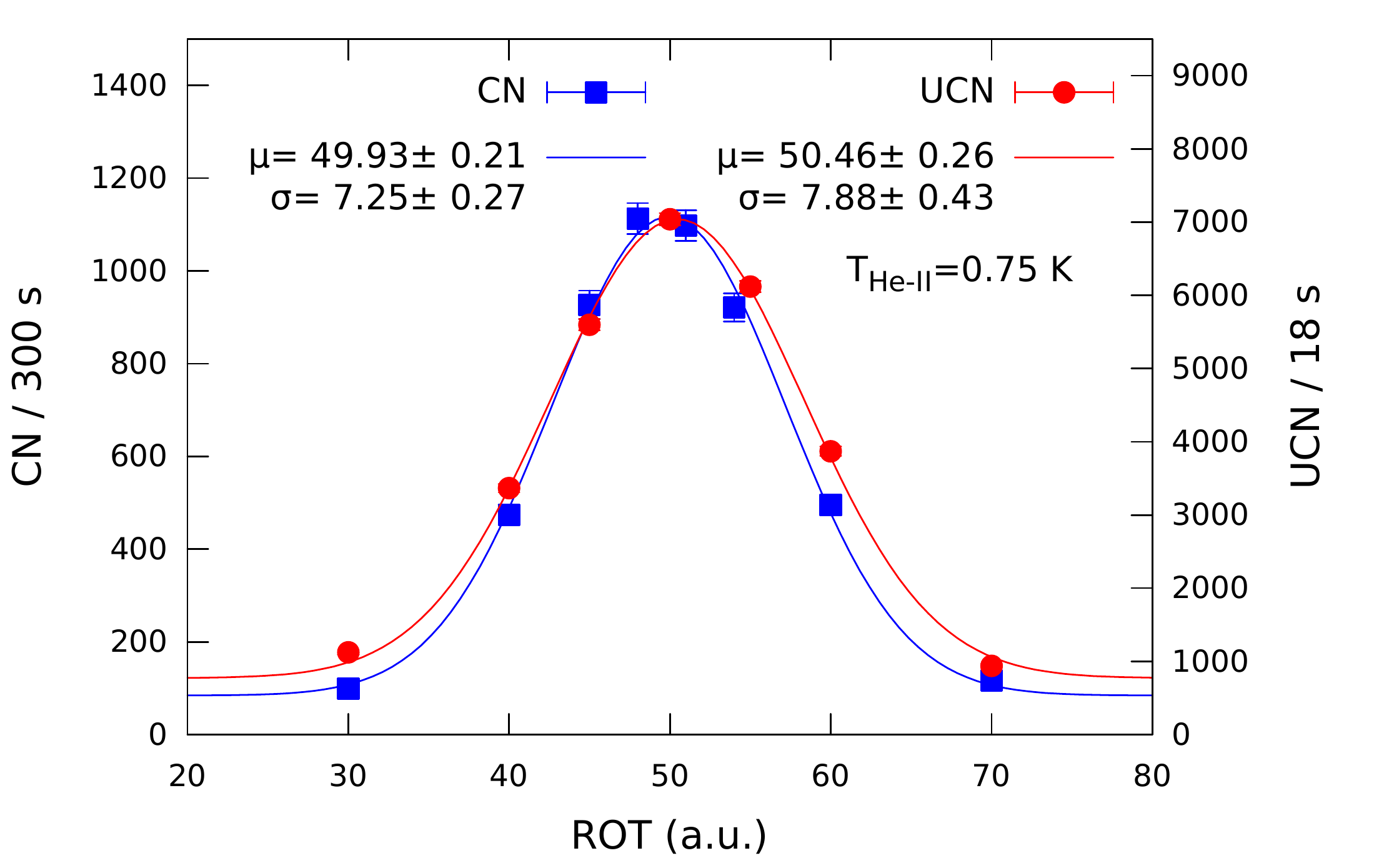}
\end{center}
\caption{Neutron rate as a function of the monochromator rotation angle, for UCNs (dots) and cold neutrons (squares).}
\label{rotscan}
\end{figure}

\label{sec:countrate}
The position and orientation of the monochromator can be adjusted remotely with five parameters: rotation, two tilt angles and two translation axes. These parameters are optimized by maximizing the UCN flux out of the source.
The most critical parameter is the rotation of the monochromator, for which the neutron countrate varies by $80$\% of maximum a few degrees away from the optimal position. For the other parameters, within their whole range, the countrate varies at most by $40$\%.
The result of the scan is shown in Fig.~\ref{rotscan}. We checked that maximum UCN flux coincides with the maximum cold neutron flux, indicating that the setup is aligned correctly.

\subsection{Differential neutron flux}

\begin{figure}

\begin{center}
\includegraphics[width=0.49\textwidth]{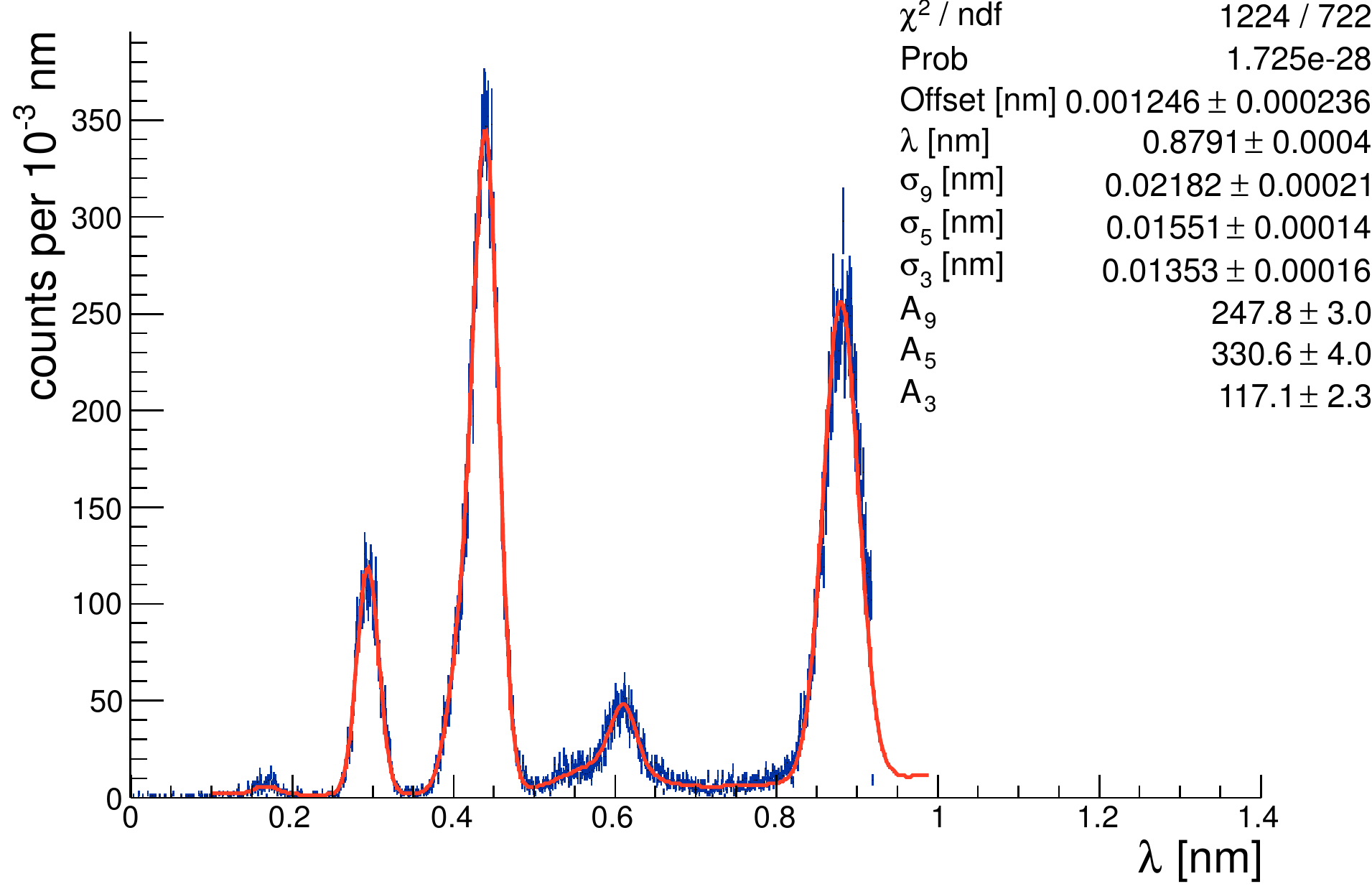}
\end{center}
\caption{Long-distance ($78$~cm) TOF spectrum. It is assumed that the three main peaks arise from the three first orders of Bragg diffraction which fixes the wavelength $\lambda/2$ and $\lambda/3$ for the second and third order peaks. There are 21 other free parameters to account for the offset of the X-axis, the amplitudes and width of the Gaussian peaks, as well as ad-hoc background description.}
\label{beam_spectrum}
\end{figure}

We have characterized the wavelength distribution in the beam using the time of flight (TOF) technique. Two measurements were performed: the first over a flight length of $78 \pm 1$~cm, the second over $28 \pm1$~cm. 
The spectrum of the first measurement is presented in Fig.~\ref{beam_spectrum}, together with a fit of the peaks. We obtained for the first order peak the central wavelength $\lambda = 0.879(11)$~nm. The uncertainty is dominated by the error on the flight length, which is itself defined by the uncertainty of knowledge of the position in the gaseous detector where the reaction occured. The width of the peak, $\sigma =0.022$~nm, is compatible with the expected TOF resolution. 

The relative intensities of the peak shown in Fig.~\ref{beam_spectrum} do not reflect the relative intensities of the corresponding neutron fluxes due to at least three important effects. First, the detector efficiency is proportional to $\lambda$ according to the $1/v$ law. Second, losses in air, estimated to be $13$~\% for $0.89$~nm neutrons, are also proportional to $\lambda$. Finally, since different wavelength components of the beam have different angular divergencies, the geometrical acceptance of the setup is also wavelength dependent. To suppress the latter, a second TOF measurement with a shorter flight length of $28 \pm 1$~cm was performed.

\begin{figure}

\begin{center}
\includegraphics[width=0.49\textwidth]{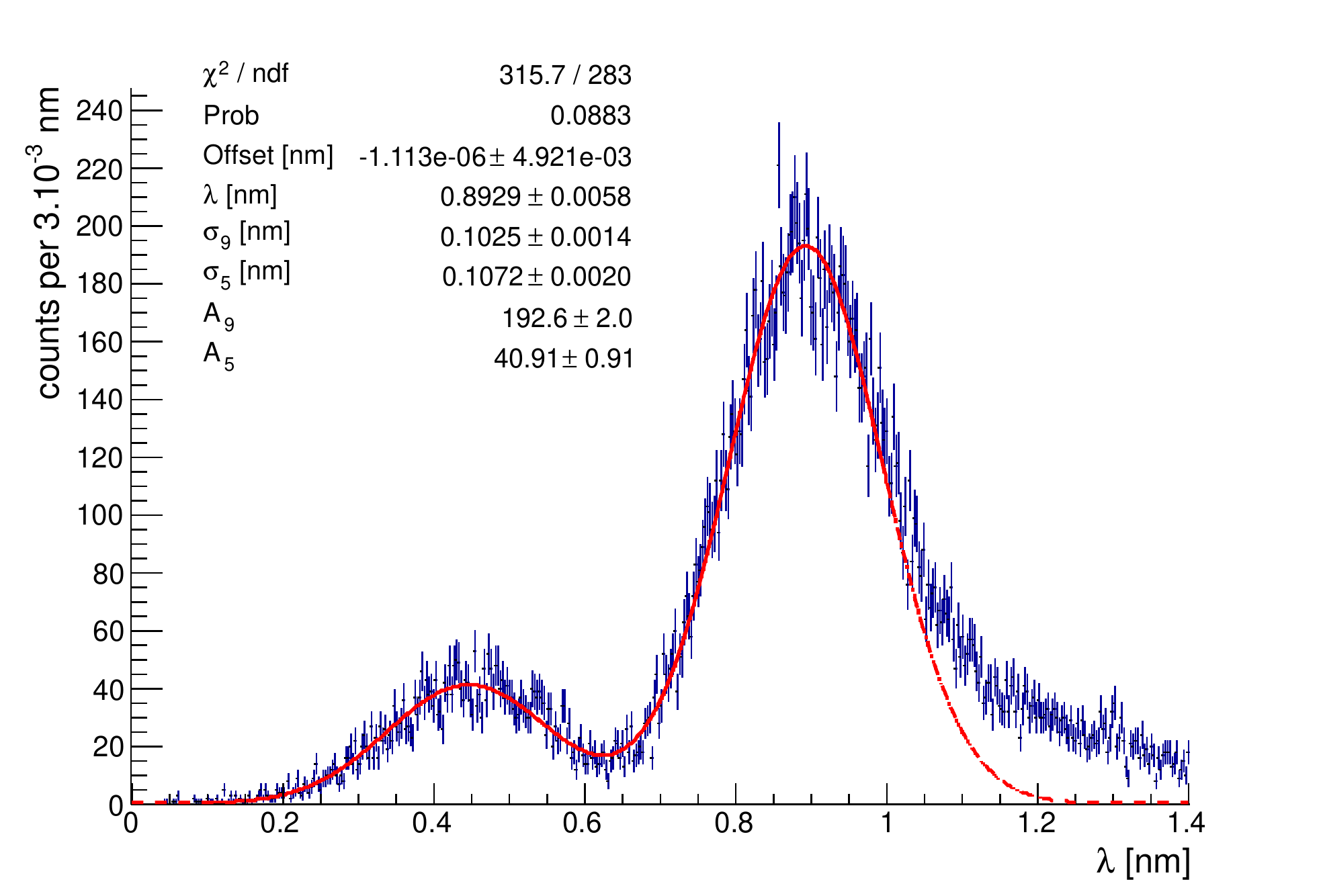}
\end{center}
\caption{Short-distance ($28$~cm) TOF spectrum. The analysis is similar to that for Fig.~\ref{beam_spectrum}. The dashed line is here to guide the eye, and the corresponding data points were not included in the fit.}
\label{tof_count}
\end{figure}

We show in Fig.~\ref{tof_count} the TOF spectrum recorded at the short distance. After correcting for the detector efficiency, we determined the relative contribution of the first and second order diffraction peaks to the total flux as
\begin{equation}
\label{eq:plambda}
p_{0.89} = 0.74 \qquad p_{0.445} = 0.26
\end{equation}

\subsection{Integral neutron flux}

To determine the integral neutron flux of the H172A beamline, we have used the standard gold foil activation technique that allows to measure the capture flux density defined by

\begin{equation}
\Phi_{\rm gold}=\int\frac{d\phi}{d\lambda}\frac{\lambda}{\lambda_{ \rm th}} d \lambda.
\end{equation}
where $\lambda_{\rm th}=0.18$~nm is the wavelength of thermal neutrons.
We have obtained in July 2012 with a reactor power of $48$~MW: 

\begin{equation}
\Phi_{\rm gold}=6.5\times 10^8\text{ cm}^{-2}\text{s}^{-1}.
\end{equation}

Knowing the proportion of each diffracted peak $p_\lambda$, neutron flux in the $0.89$ peak can be calculated:
\begin{footnotesize}
\begin{equation}
\label{eq:goldfoil}
\Phi_{[0.89\text{ nm}]}=\Phi_{\rm gold}\times\frac{0.18\text{ nm}}{0.89\text{ nm}}\times\left(\sum_{\lambda \in \rm peaks}\frac{p_{\lambda}}{p_{0.89}}\times\frac{\lambda}{0.89\text{ nm}}\right)^{-1}.
\end{equation}
\end{footnotesize}
From the differential neutron flux results (eq.~\ref{eq:plambda}), we estimate $\Phi_{[0.89\text{ nm}]}=(1.1\pm0.1)\times10^{8}\text{ cm}^{-2}\text{s}^{-1}$.
Although the intrinsic width of the peak is unknown, it cannot be larger than the measured TOF width $\sigma = 0.02$~nm. Assuming a Gaussian shape for $\frac{d\Phi}{d\lambda}$, we deduce a worst case scenario estimate (with our limit on $\sigma$) of the differential neutron flux at $0.89$~nm:
\begin{equation}
\left.\frac{d\Phi}{d\lambda}\right|_{0.89 \rm nm} =\Phi_{[0.89 \rm nm]}\times\frac{1}{\sigma\sqrt{2\pi}}.
\end{equation}

The actual flux inside the UCN conversion volume is further reduced by two effects. First, the walls of the conversion volume ($0.85$~mm thick aluminum and $1$~mm thick beryllium) attenuates the beam by a factor of $0.895$. Then, because of the angular divergence of the beam of $30$~mrad, the fraction of the cold beam interacting with the source is
$0.72$ (estimated with a Monte-Carlo simulation).

Thus, the effective $0.89$~nm differential flux inside the source is 
\begin{equation}
\left.\frac{d\Phi}{d\lambda}\right|_{0.89\text{ nm}}^{\rm eff}=(1.3\pm0.2)\times10^{9}\text{ cm}^{-2}\text{s}^{-1}\text{ nm}^{-1}.
\end{equation}
with the reactor power of $48$~MW.
As the conversion rate in BeO vessel can be estimated~\cite{SchmidtWellenburg2009267}, the volumic production rate is
\begin{equation}
P=(4.97\pm0.38)\times 10^{-9}\text{ nm} \cdot \text{cm}^{-1}\times\left.\frac{d\Phi}{d\lambda}\right|_{0.89\text{ nm}}^{\rm eff}.
\end{equation}
we deduce that we produce $P\times V=32000$ UCN/s in our $5$ l-vessel with the reactor power of $48$~MW.

\section{UCN source}

The GRANIT superthermal UCN source is an evolution of the SUN1 apparatus \cite{PhysRevLett.107.134801,Piegsa}. The UCN conversion volume consists of a vessel made out of BeO/Be filled in with superfluid $^4$He, where $0.89$~nm neutrons are down-scattered to the UCN energy range by resonant phonon excitation. This rectangular volume of the size $7 \times 7 \times 100$~cm$^3$ is placed in continuity with the neutron guide, which is also of squared section ($7 \times 7$~cm$^2$). The conversion volume is encased in a cryostat that allows to cool the volume down to below $0.8$~K. 
A first commissioning of SUN1 at its definitive position was reported in \cite{PhysRevLett.107.134801,Piegsa}, where a large diameter extraction guide from the source to a UCN detector was installed. In order to avoid diluting of UCNs in the phase-space density, a more elaborate extraction guide assembly, with smaller diameter in particular, was designed and built. In the following we report the necessary modifications to the source as well as the current performances. 

To determine these performances, a temporary set-up was installed. A box (filled in with Argon to reduce the UCN losses) was connected to the UCN extraction guide outside the clean room.
That set-up allowed us to perform our first measurements with UCNs.

\subsection{Cryostat}

\begin{figure}

\begin{center}
\includegraphics[width=0.49\textwidth]{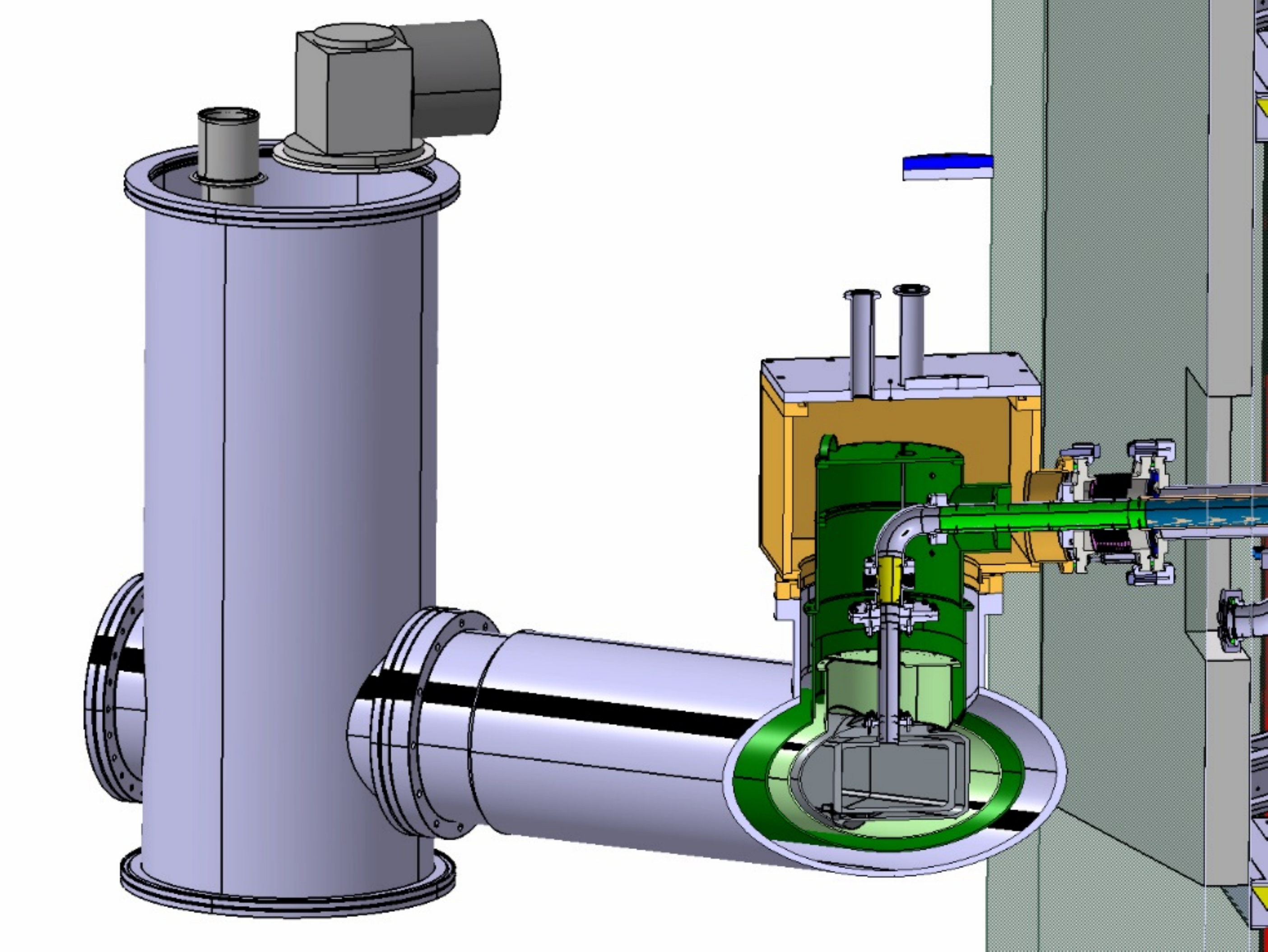}
\end{center}
\caption{Extraction guides from the source to the spectrometer. The extraction guides are composed of several tubular elements, which are thin foils of stainless steel inserted inside tubes. The design of the guides allows compensating for the misalignment between the source and the spectrometer.}
\label{extract}
\end{figure}

\begin{figure}

\begin{center}
\includegraphics[width=0.49\textwidth]{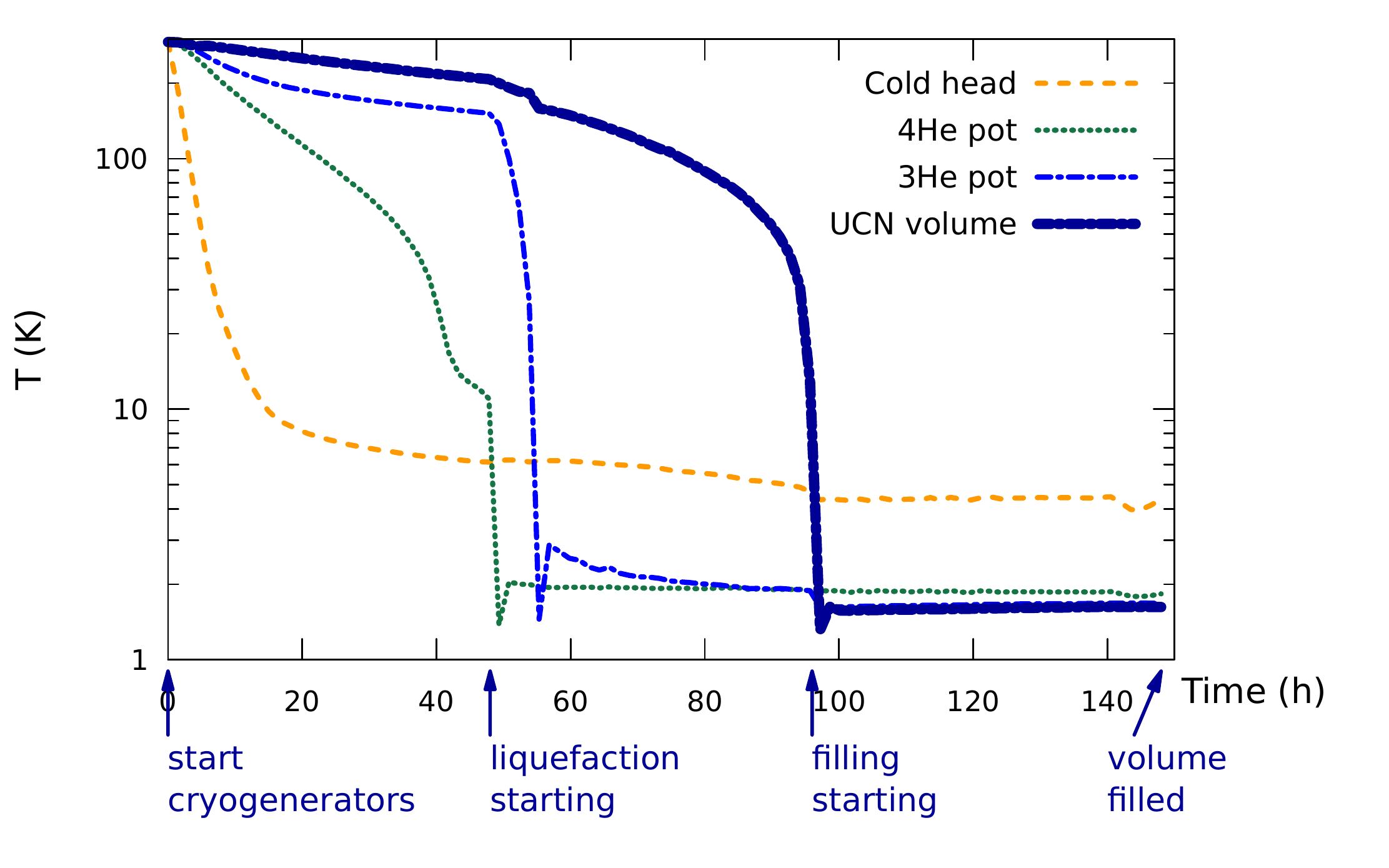}
\end{center}
\caption{A successful cooldown in $100$ hours. The temperature of several areas of the cryostat are constantly monitored. It takes $48$ hours from the beginning to start $^4$He liquefaction, that helps cooling the conversion volume. $48$ hours later, the conversion volume is cold enough for keeping liquid $^4$He. $52$ additional hours are needed to fill in this volume. Then, \mbox{He-II} can be cooled down to the temperature of $0.74$ K.}
\label{cooldown}
\end{figure}

With the assembly of the new extraction presented in Fig.~\ref{extract}, that connects the coldest part to the ambient temperature, first tests showed a loss of cryogenic power. In the first configuration it was not possible to cool down the UCN volume below the temperature of $1$~K, thus we had to increase the cryogenic power of the refrigerator. A first temporary set-up using a LN$_2$ cooling system for the thermal screens (around the inner parts of the extraction) was installed in order to validate that increasing power could solve the problem. This set-up permitted us to reach the temperature of $0.74$~K. In the second configuration this LN$_2$ cooling system was replaced with a Sumitomo cold head 150 W@77~K. The outlet box of the cryostat and some thermal screens were also replaced. The current cryogenic system is now adapted for our configuration. Fig.~\ref{cooldown} shows a cool down and filling of the conversion volume. 

\subsection{Separation windows}
\label{sec:extraction}

In several occurrences, we need to physically separate volumes at different pressures keeping UCN-transparency. This is the case for any gaseous UCN detector.
Also, the spectrometer vacuum has to be separated from the extraction vacuum to avoid re-heating in case of spectrometer openings. Any material separation is a cause of extra UCN losses due to quantum reflection and absorption.

\begin{figure}

\begin{center}
\includegraphics[width=0.49\textwidth]{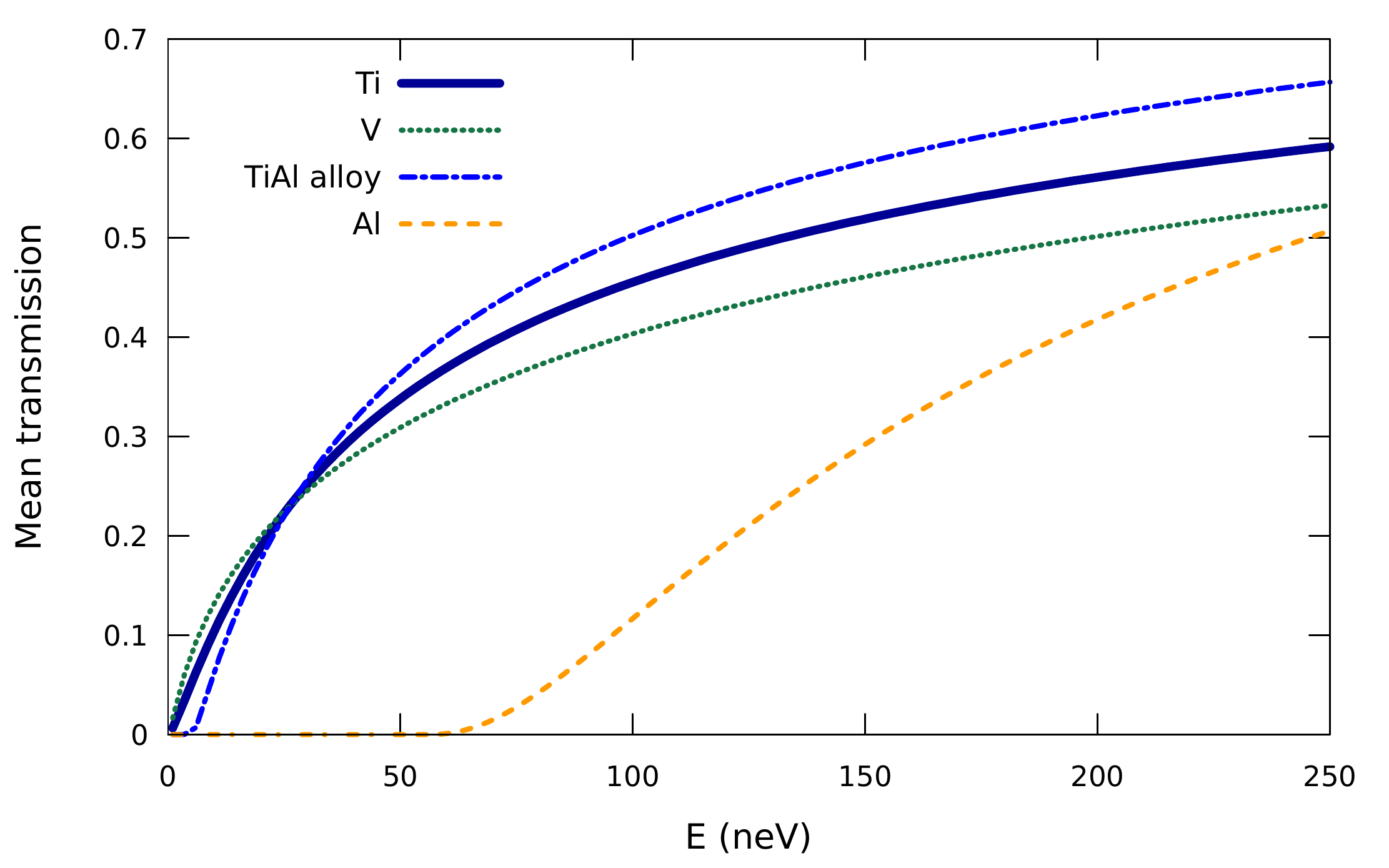}
\end{center}
\caption{An analytical calculation of the probability of transmission of UCNs through two foils ($15$~$\mu$m each for Ti, V and TiAl alloy, $30$~$\mu$m each for Al) for an isotropic UCN gas as a function of UCN energy.}
\label{transmission_isotropic}
\end{figure}

The first choice for a material was aluminum that has both low optical potential ($54$ neV, corresponding to a critical velocity of $3.2$~m/s) and small neutronic absorption cross-section ($\sim102$~barn for $5$~m/s neutrons). However, soft UCNs (with the velocity lower than $3.2$~m/s) are supposed to represent a non-negligible fraction of UCNs extractable out of the source. A simple calculation of the transmission through two windows (one in the extraction and one for the gaseous detector) considering an isotropic angular distribution of UCNs, illustrated on Fig.~\ref{transmission_isotropic}, shows that better materials exist for our application.

\begin{figure}

\begin{center}
\includegraphics[width=0.3\textwidth]{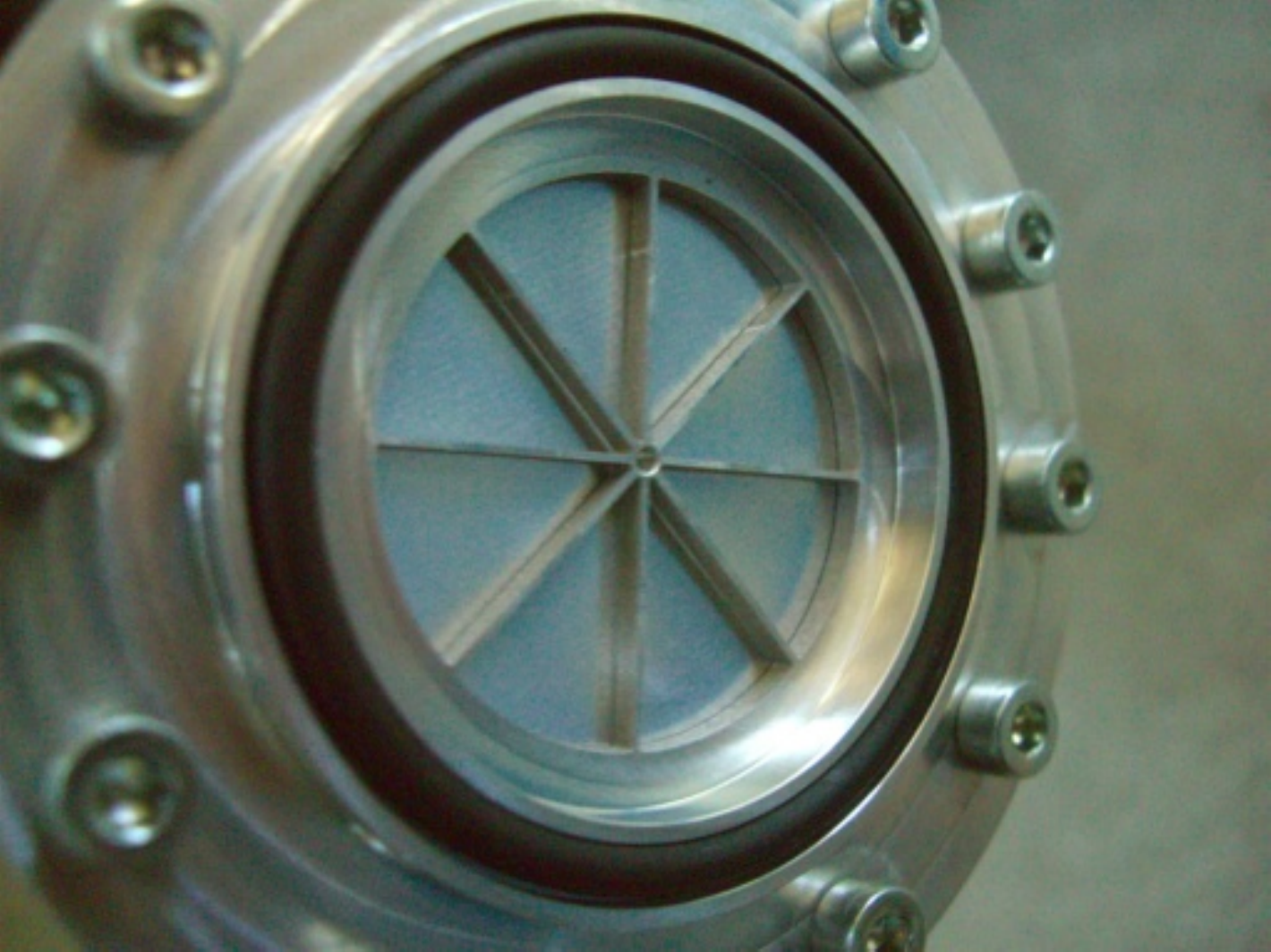}
\end{center}
\caption{Vacuum-separating window in the extraction guides. The foil must guarantee the vacuum tightness but also it should resist to an overpressure of 1.5~bar max in both directions in case of eventual fast heating of the source. Such overpressure is limited using two safety valves. The window is made with 2 aluminum half pieces milled by electro cutting reinforced by letting ribs 0.25~mm thick. Between these two parts a titanium foil 0.015~mm thick and 2 o'rings are inserted. The assembly is bolted. The window assembly had been tested with success under vacuum and under 5~bar pressure.}
\label{ribs}
\end{figure}

We conclude that titanium would be better-suited as long as soft UCNs are available, and pure titanium windows are quite easy to set-up. However, the thickness of the windows is far more critical for titanium than for aluminum (because of the absorption cross-section for neutrons). Thus, depending on mechanical constraints, the windows should be as thin as possible.
Fig.~\ref{ribs} shows the extraction window between the extraction and the spectrometer that was designed for GRANIT.

\label{sec:results}

\subsection{UCN countrate vs temperature}

\begin{figure}

\begin{center}
\includegraphics[width=0.49\textwidth]{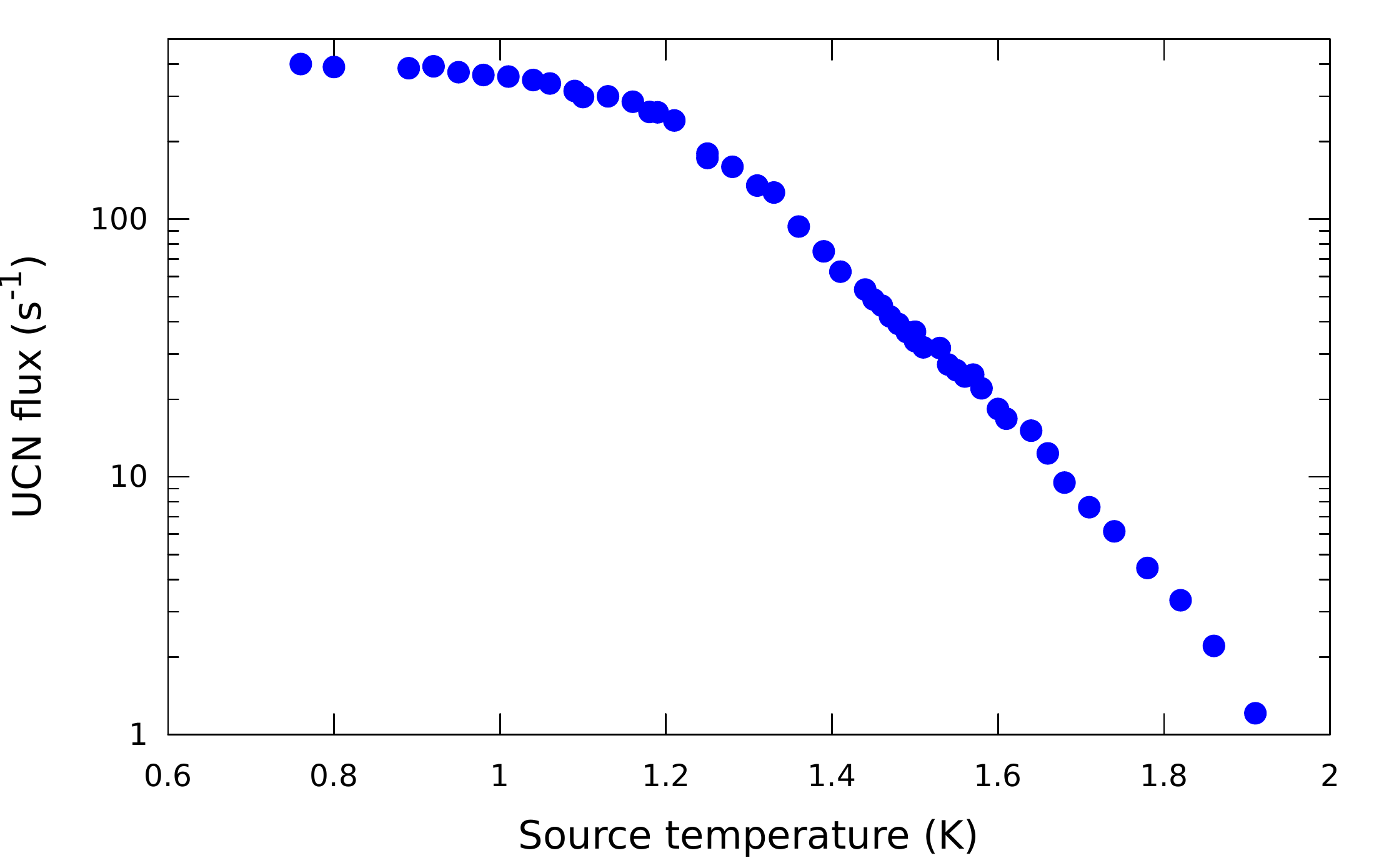}
\end{center}
\caption{UCN countrate versus the temperature of \mbox{He-II}. The cold neutron beam constantly passes through the source and the UCN valve is opened periodically.}
\label{UCNvsT}
\end{figure}

A measurement of the UCN countrate as a function of \mbox{He-II} temperature was realized with a $^3$He counter. The extraction window was made of a $30$ $\mu$m aluminum foil, as well as the detector entrance window.
The result is presented in Fig.~\ref{UCNvsT}.

Two regimes appear. When \mbox{$T\geq1.2$~K}, the UCN flux is highly dependent on temperature because of the up-scattering of UCNs in \mbox{He-II}. At lower temperature (\mbox{$T\leq0.9$~K}), the UCN flux saturates, because the main losses are caused by absorption of UCNs in the beryllium walls of the conversion volume. If this effect was an order of magnitude smaller, the transition would occur at a lower temperature and the saturation UCN density would be nearly an order of magnitude higher.

When the UCN valve of the source is open, radiative heat on \mbox{He-II} causes an important increase in temperature ($0.045$~K/min). For this reason, the valve should not be opened for longer than $10$-$15$ seconds in order to have a reliable and stable measurement, as well as to be able to cool back down \mbox{He-II} ($\sim10$ min). Thus, the source can operate in a pulsed regime at a temperature below $0.9$~K (stable as long as the opening time of the valve is short), or in a continuous mode at a higher but always stable temperature ($\sim1.3$~K), but with a smaller UCN flux.

\subsection{Source and extraction characteristic times}

We define two characteristic times for this system. The emptying time $\tau_{\rm emptying}$ is the average time necessary to extract UCNs from the apparatus. The storage time $\tau_{\rm storage}$ is the average lifetime of UCNs in the isolated source.

\begin{figure}[!b]

\begin{center}
\includegraphics[width=0.49\textwidth]{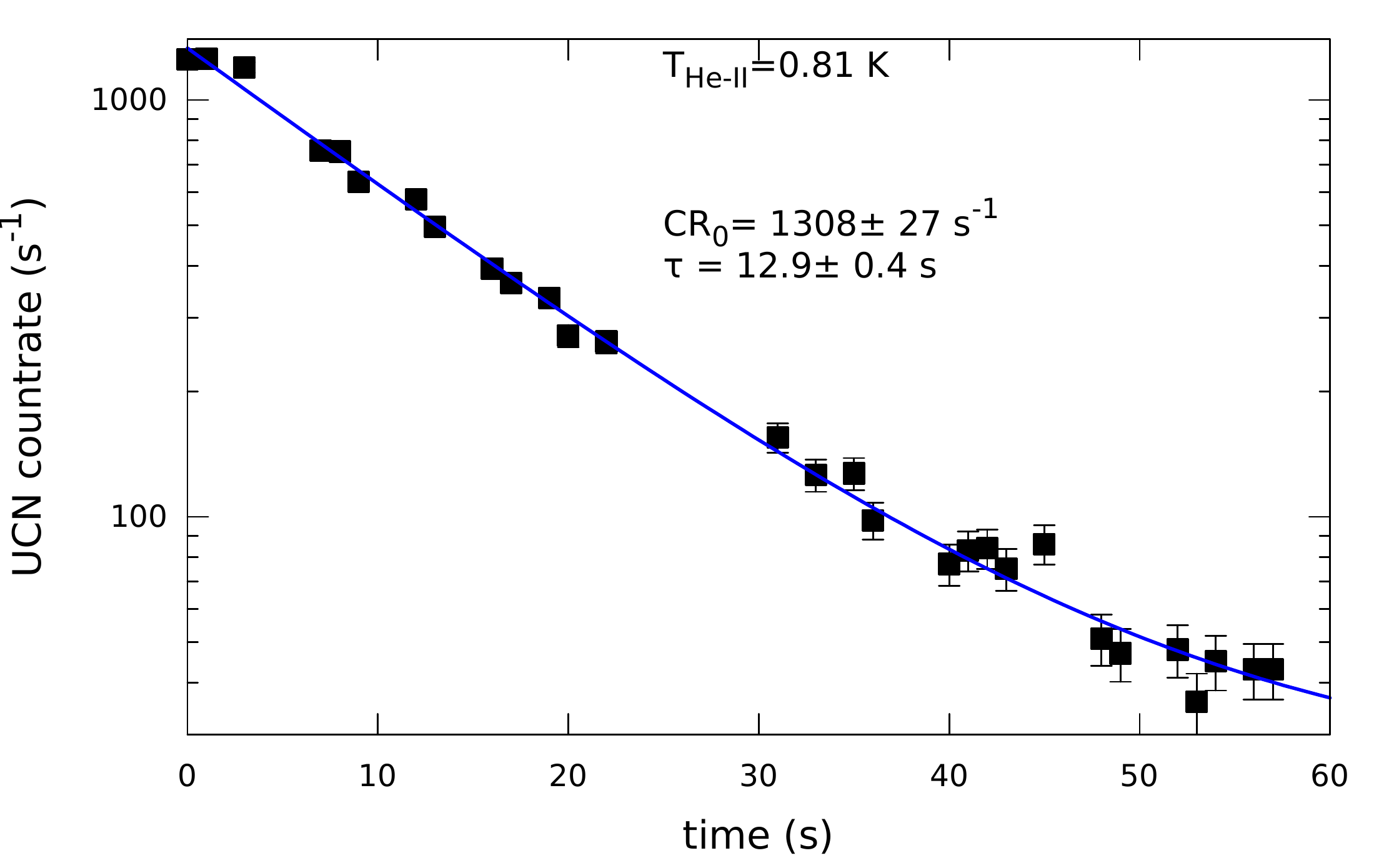}
\end{center}
\caption{Emptying time of UCNs in the source and extraction with Ti windows.}
\label{empty}
\end{figure}

\begin{table}[!b]
\caption{Emptying time of UCNs in the source and extraction with two titanium windows for several waiting times.}
\label{tab:empty}
\begin{center}
\begin{ruledtabular}
\begin{tabular}{c c}
Waiting time (s) & $\tau_{\rm emptying}$ (s)\\
\hline
$0$ & $12.9\pm0.4$\\
$50$ & $18.6\pm0.6$\\
$100$ & $23.8\pm1.4$\\
\end{tabular}
\end{ruledtabular}
\end{center}

\end{table}

The emptying time is measured by accumulating UCNs in the source during 2~min, then closing the cold beam shutter and waiting for a few seconds. The UCNs are then released through the extraction to a UCN detector, with a differential UCN count measurement. The data are fitted with a single exponential as illustrated in Fig.~\ref{empty}. The results for different wainting times are summarized in Table~\ref{tab:empty}. One can notice that the longer the waiting time, the longer the emptying time. We conclude that the softer UCNs are stored longer in the source, and that a sufficiently long time must be chosen to integrate the number of neutrons extracted without introducing a bias between measurements with different waiting times.

\begin{figure}

\begin{center}
\includegraphics[width=0.49\textwidth]{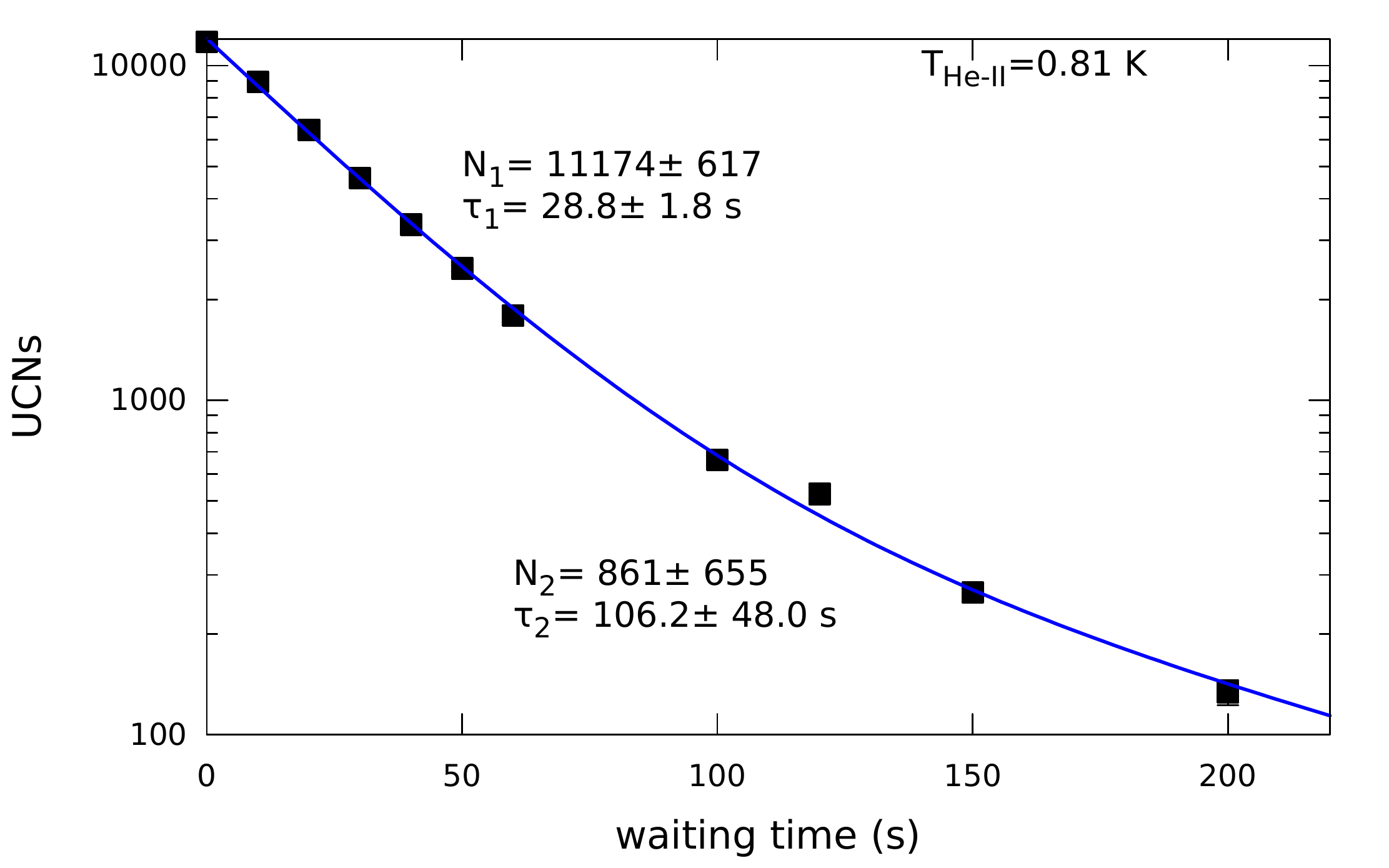}
\end{center}
\caption{Storage time of UCNs in the source with Ti windows.}
\label{storage}
\end{figure}

The storage time is measured in a similar way, but the total number of extracted neutrons is counted, and the procedure is repeated for different waiting times. This time is obtained by fitting the data with a double exponential, as shown in Fig.~\ref{storage}, and calculating the weighted geometric mean of the two decay constants. The two exponentials allow to account for several UCN populations with different velocities and storage times where a single exponential is not sufficient.

The storage time measurement was performed for several combinations of extraction and detector windows materials. The results are summarized in Table~\ref{tab:storage}.
\begin{table}
\caption{Storage time in different configurations.}
\label{tab:storage}
\begin{center}
\begin{ruledtabular}
\begin{tabular}{c c c }
Extraction & Detector & $\tau_{\rm storage}$ (s) @ $0.81$~K\\
\hline
Al & Al & $21.0\pm0.4$ \\
Ti & Al & $21.3\pm0.4$\\
Ti & Ti & $30.4\pm0.7$\\
\end{tabular}
\end{ruledtabular}
\end{center}
\end{table}
Having at least one aluminum window cuts out the soft UCNs, resulting in similar storage times for Al-Al and Ti-Al configurations. Using only titanium allows to recover the soft UCNs. The noticable increase of the storage time is expected as slower UCNs have less collisions on the source walls. The choice of titanium windows is therefore justified, and gives us access to a non-negligible fraction of the UCN velocity spectrum.

The results show that $\tau_{\rm emptying}<\tau_{\rm storage}$: the source can work in an accumulation mode where UCNs are accumulated in the source, then released all at the same time towards the spectrometer.

\subsection{UCN velocity spectrum}
\label{sec:spectrum}

\begin{figure}[h]

\begin{center}
\includegraphics[width=0.49\textwidth]{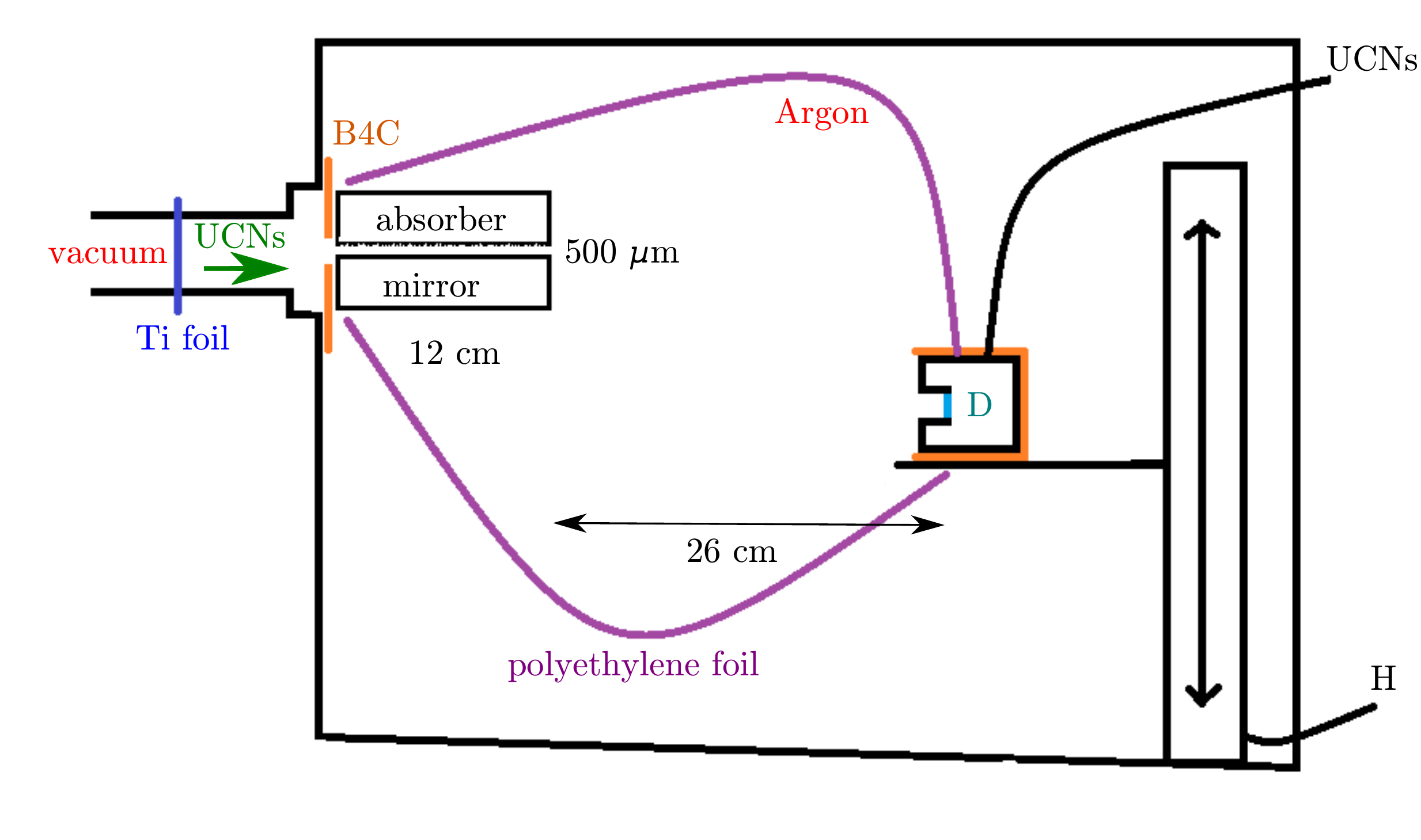}
\end{center}
\caption{Set-up for free fall measurement: in an argon-filled box, the UCN beam is collimated with an $12$~cm long absorber-mirror system forming a slit of height $500$~$\mu$m. A detector with a $15$~$\mu$m-thick titanium window is placed at the distance of $26$~cm from the slit and record the neutron count as a function of the fixed free fall height.}
\label{measurement}
\end{figure}

The velocity distribution of UCNs can be determined with a free fall experiment, as described in Fig.~\ref{measurement}. The height of free fall for an UCN with a defined horizontal velocity is:

\begin{equation}
h=\dfrac{g}{2}\left(\dfrac{d_{\rm freefall}}{v_{\rm UCN}}\right)^{2}.
\end{equation}

Assuming we have spatially isotropic distribution of UCNs within our collimation system (this assumption is valid with high accuracy for any broad angular distribution of UCNs in the extraction system), the measurement of horizontal velocity of the collimated UCNs is representative of the velocity distribution of the UCNs in the extraction.

\begin{figure}

\begin{center}
\includegraphics[width=0.49\textwidth]{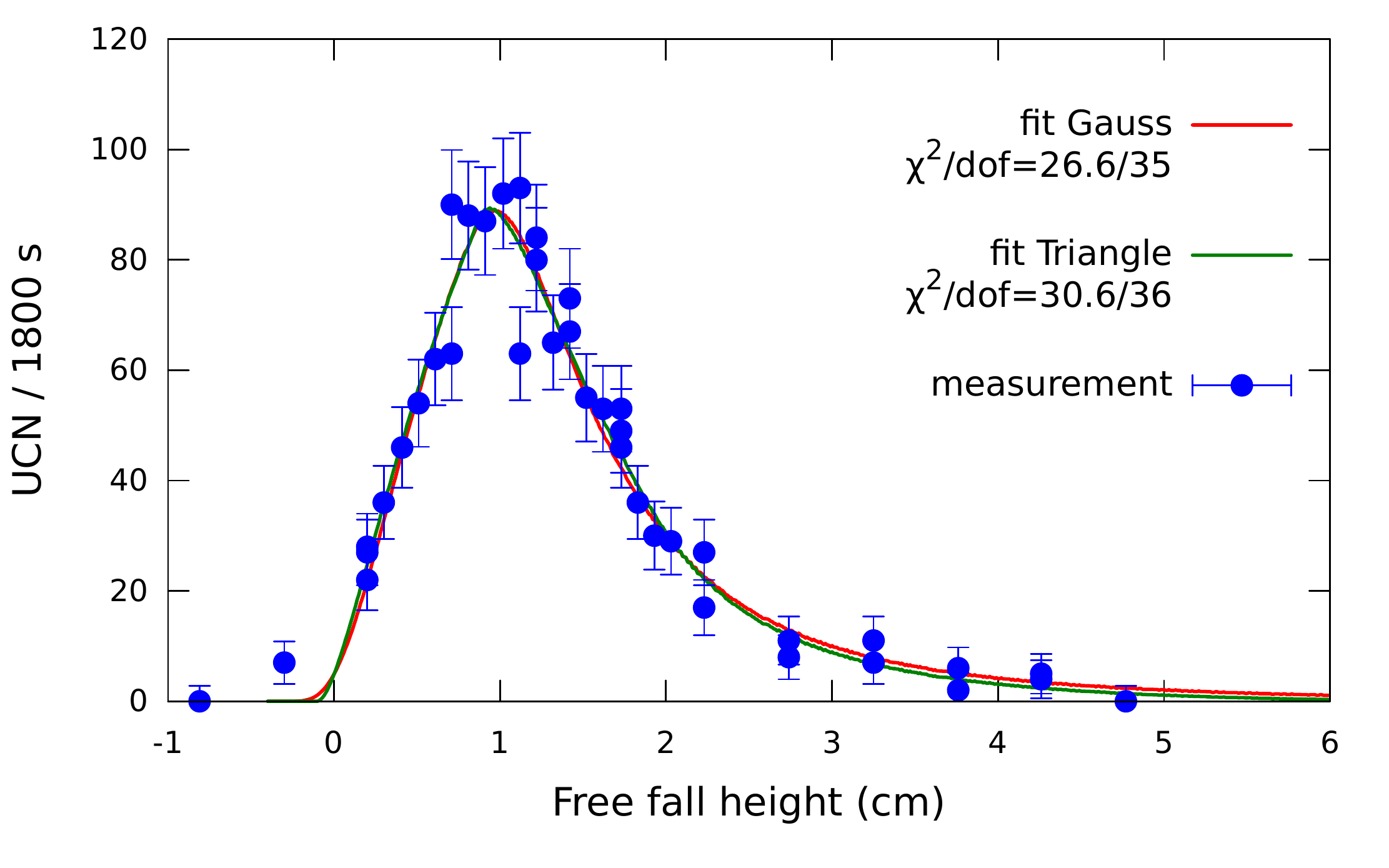}
\end{center}
\caption{Number of UCNs detected for 30 min versus the free fall height. During the measurement, the temperature of \mbox{He-II} was $1.35$~K. Since the UCN spectrum is defined by the storage and transmission properties of the whole system, and the temperature of \mbox{He-II} affects only the total count rates, the exact value of the temperature is of no importance. This measurement was fitted with a Monte-Carlo simulation, assuming an initial triangular or gaussian spectrum.}
\label{freefall_results}
\end{figure}

The recorded UCNs counts as a function of the free fall height are shown in Fig.~\ref{freefall_results}.
In order to obtain a velocity spectrum for this measurement, we make a Monte-Carlo simulation of the experiment to fit the initial velocity spectra to the data.
Two shapes were assumed for the initial spectrum: a Gaussian distribution and an asymetric triangular distribution. Both fitted correctly the data, and the obtained results for the mean and RMS of the distribution are:
\begin{equation*}
\mu_{\rm Gauss}=5.1~\textrm{m.s}^{-1}\qquad\sigma_{\rm Gauss}=1.6~\textrm{m.s}^{-1}
\end{equation*}
\begin{equation*}
\mu_{\rm Triangle}=5.3~\textrm{m.s}^{-1}\qquad\sigma_{\rm Triangle}=1.4~\textrm{m.s}^{-1}
\end{equation*}
The asymetric triangular distribution is zero below $2.2$~m/s, maximum at $4.5$~m/s and zero above $9.0$~m/s.
In both cases, we obtain a coarse UCN velocity spectrum and notice a quite wide distribution around the mean value.

Because of the low statistics (some hundreds of UCNs counted for $30$ minutes), the valve was always open (thus the temperature was $1.35$~K). In this configuration, there is no accumulation of soft UCNs in the source, thus a higher mean velocity than what we could have.
This method of measurement for the velocity spectrum of UCNs, though lacking in precision, was successful. The precision could be improved by designing a UCN detector with a dedicated geometry or a large position-sensitive detector, and a vacuum-tight environment.
The result itself is in agreement with what we expected, and confirms that the source/extraction system we use is well-suited for the GRANIT spectrometer.

\section{Spectrometer}

\subsection{Description}
\label{sec:optics}
\begin{figure}

\begin{center}
\includegraphics[width=0.49\textwidth]{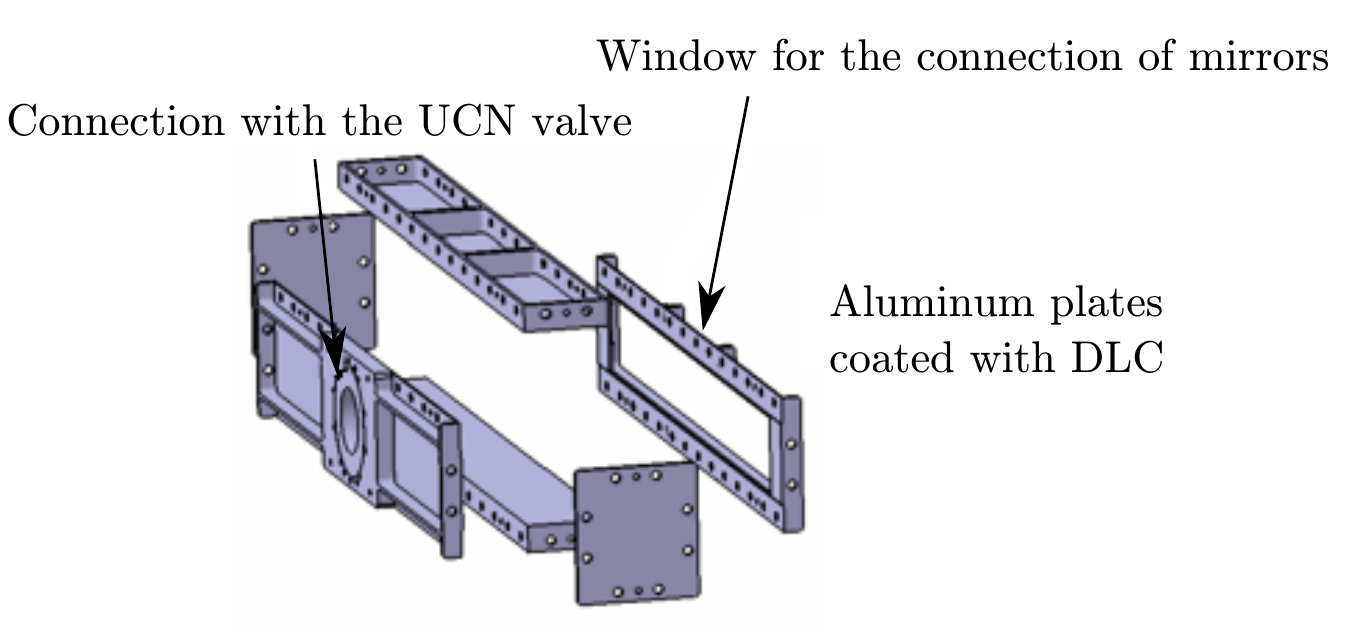}
\end{center}
\caption{The DLC-coated storage volume. The inner sizes are $40\times 40\times 340$~mm$^3$.}
\label{storage_volume}
\end{figure}

\begin{figure}

\begin{center}
\includegraphics[width=0.49\textwidth]{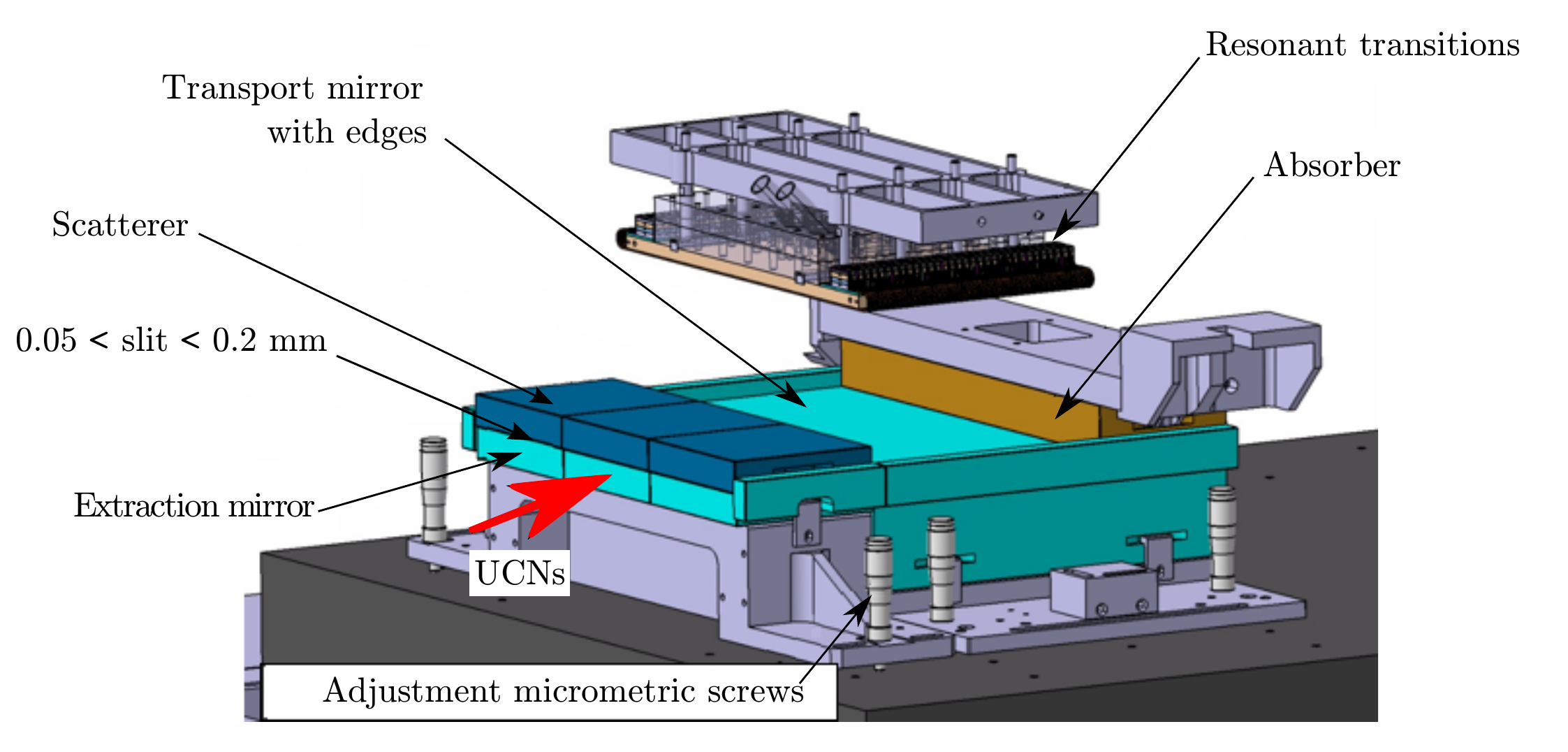}
\end{center}
\caption{Schematics and pictures of the optical elements. The extraction mirror and the scatterer have the same sizes ($300\times100\times20$~mm$^3$) and each consists of 3 pieces of $100 \times 100 \times 20$~mm$^3$ glued together. The mean roughness of the extraction mirror (floor) is very low ($0.5$~nm) to allow for specular bouncing, whereas the mean roughness of the scatterer (ceiling) is intentionally high ($5.6$~$\mu$m) to allow UCNs with too high vertical velocity to be diffused back. The transport mirror ($300\times250\times70$~mm$^3$) has a surface mean roughness of $0.5$~nm and a planarity of 80~nm. The absorber mirror is $300\times90\times30$~mm$^3$.}
\label{optics}
\end{figure}

\begin{figure}

\begin{center}
\includegraphics[width=0.49\textwidth]{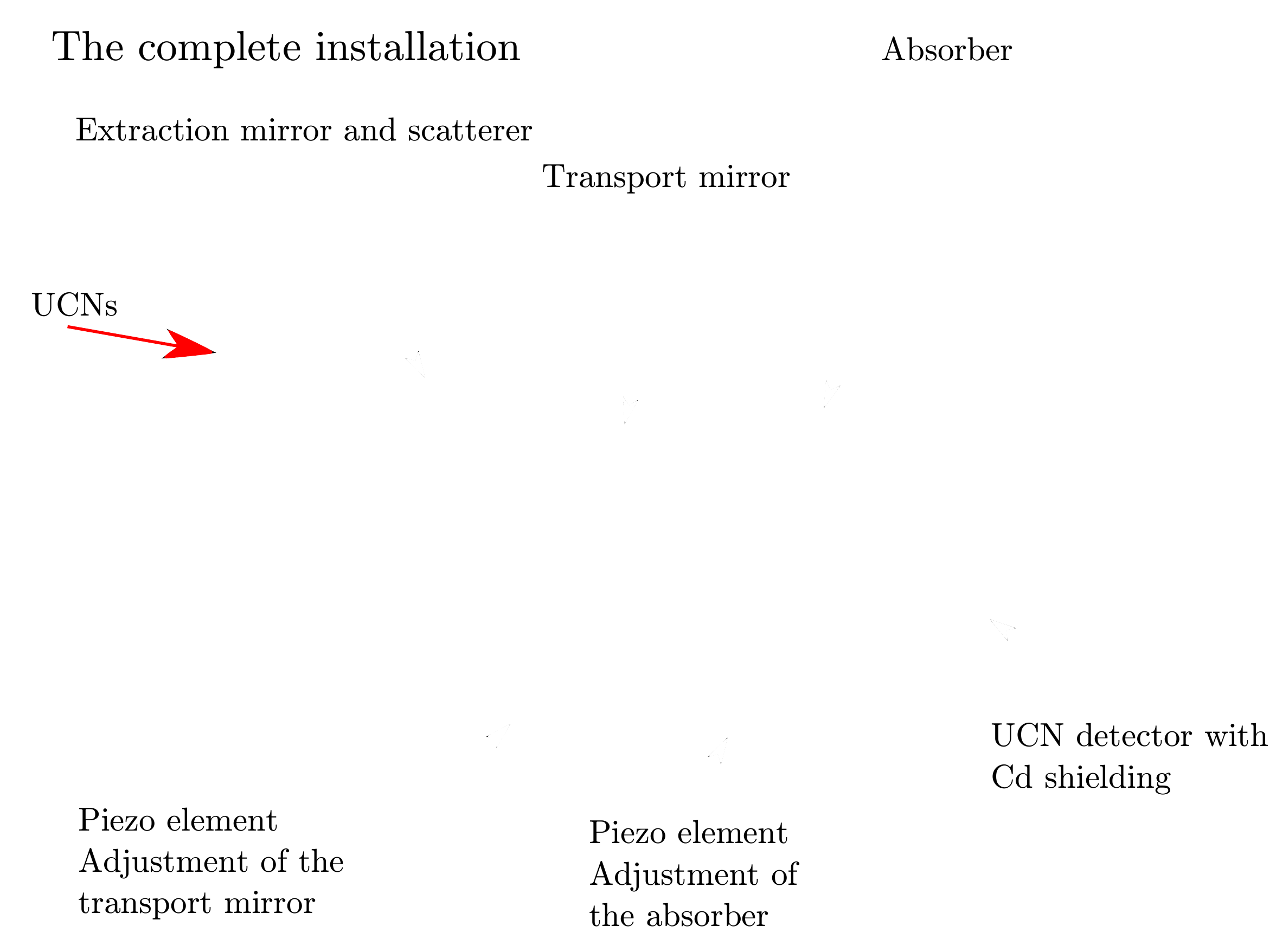}
\end{center}
\caption{The optical elements on the granit table. The extraction mirrors assembly and the transport mirror are placed on two separate adjustable supports. Their adjustment could be done with $3+3$ micrometric screws. To adjust the height and the orientation of the surface of the transport mirror with a great accuracy, we use 3 piezo-electric elements. The distance between the absorber and the transport mirror is adjustable as well using 3 piezo-electric elements. The piezos are driven from the control computer with a Labview application.}
\label{mirrors_4}
\end{figure}

UCNs are transported through the extraction guide from the He-II source to the spectrometer as shown in Fig.~\ref{extract}. The extraction guides and the extraction window are the same as the previous configuration.
The guides connect the source to an intermediate storage volume depicted in Fig.~\ref{storage_volume}. Optical elements, at the heart of the spectrometer, are connected at the exit of the storage volume. They are sets of silica pieces with different coatings and roughness states, providing different conditions to the transport of UCNs. They are described in Fig.~\ref{optics} and~\ref{mirrors_4}.
All these parts have been provided by the SESO Company in Aix en Provence (France), and Diamond-Like Carbon (DLC) coatings were applied by Advanced Material Laboratory (IN2P3, France).

The intermediate storage volume, made of several aluminum parts coated with DLC, is designed to randomize the UCNs trajectories. In order to close the storage volume, a nickel-coated butterfly valve is used, preventing the UCNs from returning into the source. The exit of the intermediate storage volume is closed by the first optical elements of the spectrometer, forming a slit and allowing only UCNs with negligible vertical velocity to be transmitted. The slit is composed of an extraction mirror and a scatterer. The scatterer is placed above the extraction mirror at a height between $50$ and $200$~$\mu$m. Both are coated with DLC in order to provide reflection of UCNs with a broad velocity range from the surfaces and thus to assure proper operation of the so-called semi-diffusive slit~\cite{Barnard2008431,SchmidtWellenburg2007623}.

\begin{figure}

\begin{center}
\includegraphics[width=0.15\textwidth]{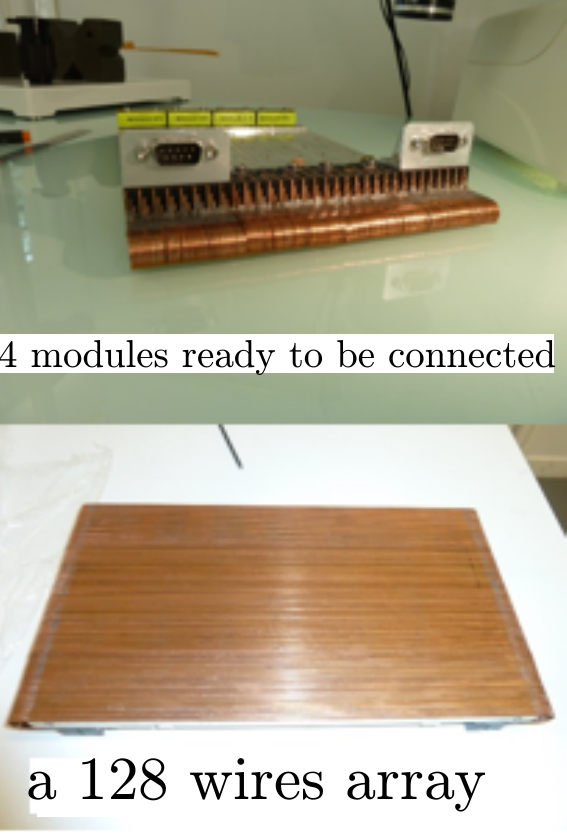}
\includegraphics[width=0.30\textwidth]{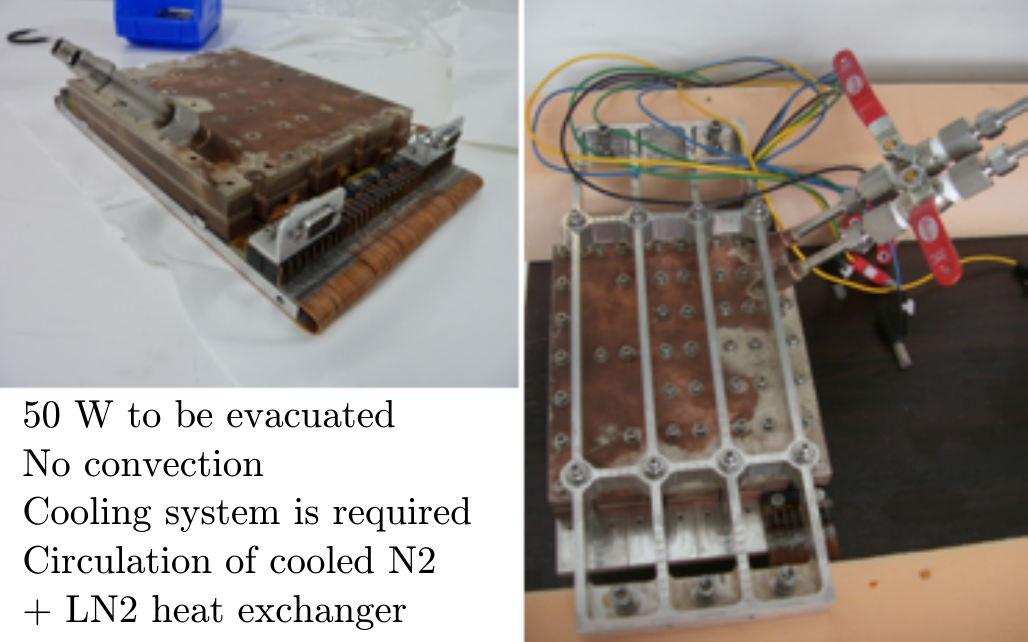}
\end{center}
\caption{The magnetic excitation array is made of 4 modules of 32 wires each ($1$~mm$^2$) constituting a 128-wire array. The wires are spaced by $0.25$~mm.}
\label{tr_coils}
\end{figure}

The second part of the spectrometer is a transport mirror, aligned with the slit (or positionned a few micrometers above/below the slit, allowing the selection of quantum states). The magnetic excitation array~\cite{PignolTR}, described in Fig.~\ref{tr_coils}, can be placed above it. To induce resonant transitions between quantum states, a periodic magnetic field gradient will be produced with an array of wires located above the transport mirror.
The third part of the spectrometer is an absorber, placed above and at the end of the transport mirror at an adjustable height to filter the quantum states. 

\subsection{Cooling of the magnetic excitation array}

The magnetic excitation array was built and tested before set-up. The current needed (5~A in average) to generate a sufficient magnetic field gradient produces 50~W of power that has to be dissipated without affecting too much the transport mirror. The wire system must be cooled down. Circulation of cooled gaseous N$_2$ will be used for that purpose.
To test the cooling in conditions as close as in the experimental set-up, the magnetic excitation array was placed in a vacuum chamber and connected to a power supply (5~A, 8~V). The heat exchanger was placed in a LN$_2$ tank at a distance of 9~m from the magnetic excitation array, and a circulating N$_2$ circuit linked the heat exchanger to the magnetic excitation array.
Results of that test were very good, and we could easily stabilize the temperature of the wires at $15$~$^{\rm o}$C during $12$~h without human intervention. The magnetic excitation array equipped with the cooling system will be installed during autumn 2014.

\subsection{Preliminary results}
\label{sec:first_setup}

\begin{figure}

\begin{center}
\includegraphics[width=0.49\textwidth]{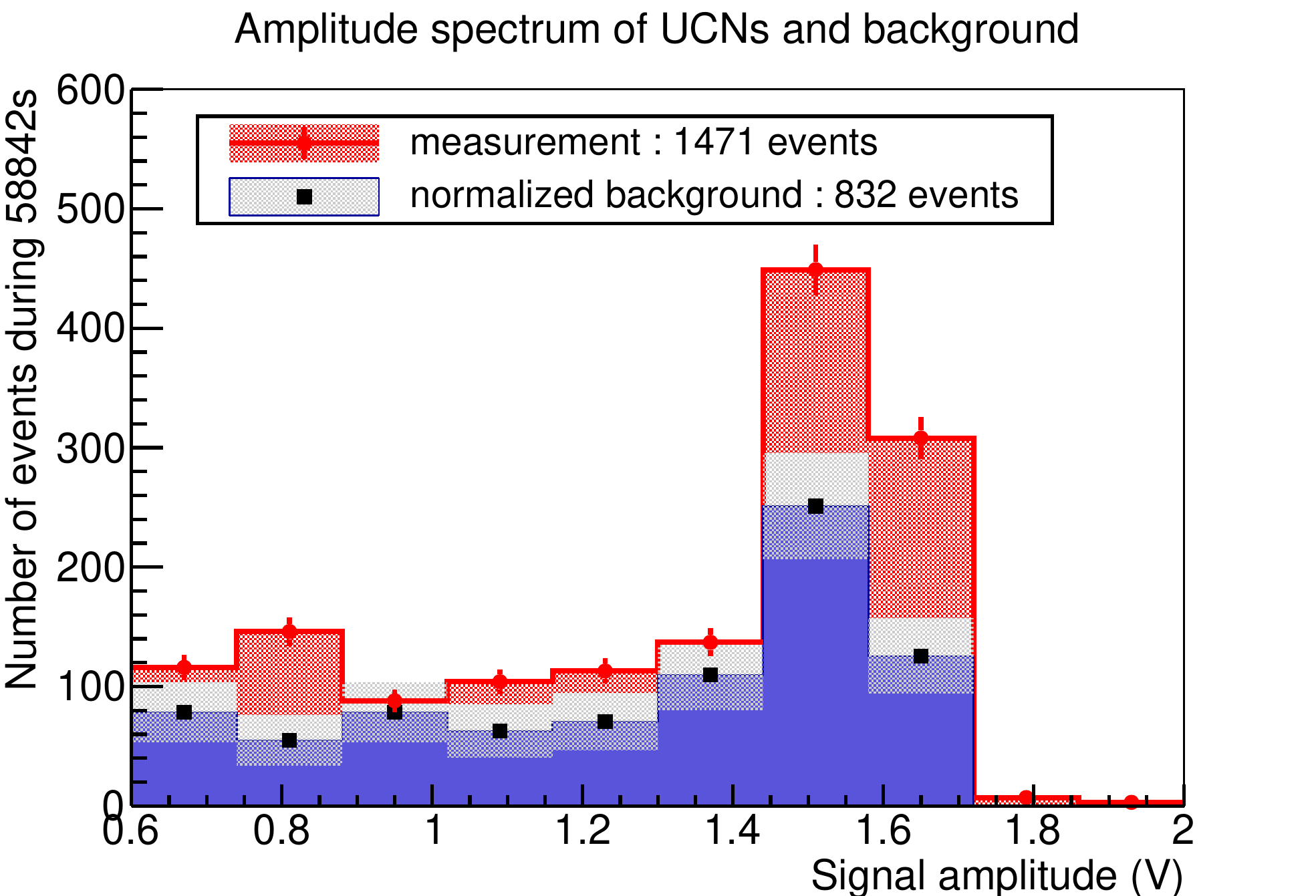}
\end{center}
\caption{Amplitude spectrum of UCNs and background. The temperature of the source was $1.35$~K}
\label{UCN_spectro}
\end{figure}

In summer 2013, we conducted the first tests of the full extraction chain. During this cycle, the source temperature could not be cooled lower than $1.35$~K.
Removing the transport mirror and the absorber, and placing the detector at the exit of the extraction slit of height $127$~$\mu$m, we measured the first UCNs in the GRANIT spectrometer.
The $^3$He detector was equipped with a titanium window.
In this configuration, we measured \mbox{$(10.9\pm1.5)\times10^{-3}$~UCN/s} as shown in Fig.~\ref{UCN_spectro}.

This countrate is one order of magnitude lower than expected in this configuration, and two orders of magnitude lower than targeted. Potential explanations include (i) a defect in the extraction which, because of the modifications of the configuration, did not exist in the previous measurements, (ii) defects in the DLC-coated storage volume, which has not been characterized on its own due to its particular geometry and (iii) the collection of impurities on a cold spot on the extraction window. This opens the way to further improvements and tests in the next cycles, which could not be done immediately because of the long reactor shutdown right after this measurement.
In addition, the background, estimated to $14\times10^{-3}$~events/s, is high compared to the signal because of background fast neutrons in the reactor building. Extra-shielding will be added to the installation in order to reduce this background.

\section{Simulation}

STARucn (Simulation of Transmission, Absorption and Reflection of ultracold neutrons) is a public Monte-Carlo software designed to simulate experimental set-ups and guides for UCNs, developped at LPSC Grenoble~\cite{starucn}.
It relies heavily on CERN's ROOT packages~\cite{root}. Its main features are modularity, easy configuration of geometry and simulation, propagation of UCNs, with or without gravity, interaction in volumes through effective lifetime, interaction at surface (quantum reflection or transmission, surface losses due to absorbtion upon reflection, specular and diffuse reflection).
These features had been successfully benchmarked against analytical calculations and existing codes such as MCUCN of PSI and GEANT4UCN~\cite{geant4ucn}.

In STARucn, a material is described with its Fermi potential $V_F$, its fraction of diffuse reflections $d$, its loss factor $\eta$ defined as the ratio of imaginary to real part of the Fermi potential and its effective lifetime $\tau$ in the material.

\begin{figure}

\begin{center}
\includegraphics[width=0.49\textwidth]{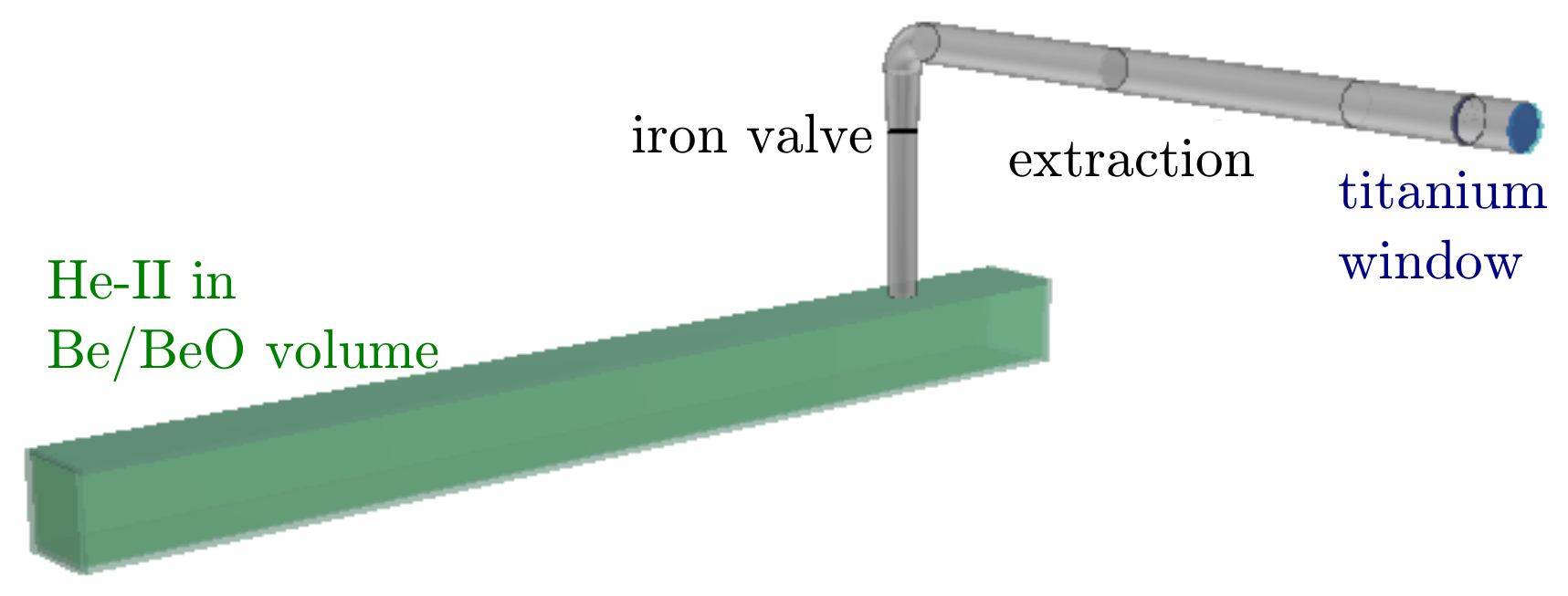}
\end{center}
\caption{Geometry described and used in the simulation.}
\label{geom_vezzu}
\end{figure}

Several measurements of the source characteristics can be used to estimate the unknown values of the parameters of the various materials.
Our simulation (with geometries such as in Fig.~\ref{geom_vezzu}) shows that the build-up time within the source is only driven by the losses on the Be and BeO surfaces of the source.
Using results from 2010 measurement~\cite{PhysRevLett.107.134801,Piegsa}, we set $\eta_{\rm Be}=2.7\times 10^{-4}$.
In a similar way, the emptying time depends mostly on the reflective properties of the stainless steel extraction. Assuming this property is uniform in all the extraction, the best fit is found at $d_{\rm steel\ extraction}=3\%$. The parameter $d_{\rm Be}$ was arbitrarily set to $0.1$ because the simulation showed that it had no significant influence on any measurement.

Assuming a production of $38000$ UCN/s isotropically distributed in phase space in the source (corresponding to the reactor power of $56$~MW), the expected number of UCNs detected in the simulation is around $5$ times higher than what is measured. No satisfactory explanation was found: for instance, a $12$~cm$^2$ hole in the extraction combined with a high $\eta_{\rm steel\ extraction}$ would be needed. The most relevant figures are in Table~\ref{tab:starucn}.

\begin{table}
\caption{Comparison of previous and recent measurements with simulations. The number of extractible UCNs correponds to the number of UCNs available in the source after the UCN density has been saturated according to the simulation. The number of extracted UCNs corresponds to the number of UCNs detected during the experiments. The measured (resp. simulated) extraction efficiency is the ratio of the number of UCNs detected during the measurement (resp. the simulation) to the simulated number of extractible UCNs. The missing factor is calculated according to the discrepancy between these two efficiencies.}
\label{tab:starucn}
\begin{ruledtabular}
\begin{tabular}{c c c}
Configuration & 2010~\cite{PhysRevLett.107.134801,Piegsa} & 2013 \\
\hline
Storage time (meas.) & 67 s & 30 s\\
Extractible UCNs (sim.) & $2.2\times10^6$ & $1.0\times10^6$\\
Extracted UCNs (meas.) & $274000$ & $21000$ \\
Extraction efficiency (meas.) & $13$\% & $2.1$\%\\
Extraction efficiency (sim.) & $65-50$\% & $15-5$\%\\
Missing factor & $4-5$ & $2.5-7$ \\
\end{tabular}
\end{ruledtabular}

\end{table}

Possible candidates to explain these losses are a larger angular divergency of the incoming beam of cold neutrons in the source than estimated, an eventual misalignement of the monochromator reducing the production rate, eventual defects or losses in the extraction and/or degradation of the BeO source volume. These options will be investigated during the next ILL reactor cycles.

Moreover, simulating the experiment of section~\ref{sec:first_setup}, with the corresponding countrate as measured in~\ref{sec:countrate} and the spectrum as measured in~\ref{sec:spectrum}, the number of neutrons measured is $10$ times lower than simulated. However, no other measurement during this cycle can rule out a problem with the source or the extraction.

\section{Improvements}

These tests of all elements of the GRANIT experiment prompt us to modify some parts of the set-up. Some of improvements are quite easy to perform. The cleanliness of the extraction guides could be easily improved, starting with a rigorous and defined production process, then having good procedures for washing, stoving and packaging each parts just before installation.

Moreover, due to the narrow UCN velocity range produced in the UCN source, DLC coating in extraction, mirrors, and intermediate storage volume is no longer crucial. For the next steps, we will replace the stainless steel foils of the extraction (that were dued to be DLC-coated) by electro-polished tubes. This will decrease the UCN losses due to the properties of the surface (mainly roughness and hydrogen adsorption) of these foils and will considerably help for the assembly of the guide, as well for connecting source to spectrometer, thus diminishing risks of damage.

For same reasons the intermediate storage volume will be replaced with a new one made of Oxygen free high conductivity copper and in a cylindrical shape thus increasing the life time of UCNs in the volume and decreasing UCN losses.
Another effort must be done to adjust the optical elements with more reliability without the contact of a probe. A solution using laser sensors is under study.

\section{Conclusion}

We have overviewed the current status of the GRANIT facility. First complete test of the GRANIT UCN source and spectrometer was performed during the last reactor cycle in 2013. Further improvements are identified basing on the measured results, and they are being implemented.

\section{Conflict of Interests}

The author declares that there is no conflict of interests
regarding the publication of this paper.

\section{Acknowledgments}
We thank all the members of the GRANIT collaboration.



\end{document}